\documentclass[draftclsnofoot, 12pt,onecolumn]{IEEEtran}
\usepackage{epsfig,graphicx,subfigure,psfrag,amsmath,cases,bm}
\usepackage{latexsym,amssymb,algorithm,mathtools}
\usepackage{algorithmic}
\usepackage{color}
\usepackage{url}
\usepackage{scrtime}
\usepackage{stfloats}
\usepackage{tablefootnote}
\usepackage{cite}
\usepackage{amsfonts,mathabx}
\usepackage{textcomp}
\usepackage{xcolor,capt-of}
\usepackage{multirow}
\usepackage{geometry}
\allowdisplaybreaks

\author{Dongfang Xu, Vahid Jamali, Xianghao Yu, Derrick Wing Kwan Ng, and Robert Schober
\thanks{This work was presented in part at the IEEE Wireless Communications and Networking Conference (WCNC) 2021 \cite{xu2021resource}.}
\thanks{D. Xu and R. Schober  are with the Institute for Digital Communications, Friedrich-Alexander-University Erlangen-N\"urnberg, 91054 Erlangen, Germany (e-mail: {dongfang.xu, robert.schober}@fau.de).}
\thanks{V. Jamali is with the Department of Electrical Engineering, Princeton University, Princeton, NJ 08544 USA (e-mail: jamali@princeton.edu).}
\thanks{X. Yu is with Department of Electronic and Computer Engineering, The Hong Kong University of Science and Technology, Hong Kong (e-mail: eexyu@ust.hk).}
\thanks{D. W. K. Ng  is with the School of Electrical Engineering and Telecommunications, University of New South Wales, Sydney,
NSW 2052, Australia (e-mail: w.k.ng@unsw.edu.au).}
}

\newtheorem{T-Prob}{Transformed Problem}

\DeclareMathOperator{\mino}{minimize}

\newcommand{\qed}{\hfill \ensuremath{\blacksquare}}
\newtheorem{Remark}{Remark}

\title{Optimal Resource Allocation Design for Large IRS-Assisted SWIPT Systems: A Scalable Optimization Framework}

\begin{document}
\maketitle
\vspace*{-10mm}
\begin{abstract}
%\vspace*{-4mm}
In this paper, we study the optimal resource allocation algorithm design for large intelligent reflecting surface (IRS)-assisted simultaneous wireless information and power transfer (SWIPT) systems. To facilitate efficient system design for large IRSs, instead of jointly optimizing all the IRS elements, we partition the IRS into several tiles and employ a scalable optimization framework comprising an offline design stage and an online optimization stage. In the offline stage, the IRS elements of each tile are jointly designed to support a set of different phase shift configurations, referred to as transmission modes, while the best transmission mode is selected from the set for each tile in the online stage. Given a transmission mode set, we aim to minimize the total base station (BS) transmit power by jointly optimizing the beamforming and the transmission mode selection policy taking into account the quality-of-service requirements of information decoding and non-linear energy harvesting receivers, respectively. Although the resource allocation algorithm design is formulated as a non-convex combinatorial optimization problem, we solve it optimally by applying the branch-and-bound (BnB) approach which entails a high computational complexity. To strike a balance between optimality and computational complexity, we also develop an efficient suboptimal algorithm capitalizing on the penalty method and successive convex approximation. Our simulation results show that the proposed designs enable considerable power savings compared to several baseline schemes. Moreover, our results reveal that by properly adjusting the numbers of tiles and transmission modes, the proposed scalable optimization framework indeed facilitates online design for large IRSs. Besides, our results confirm that the advocated physics-based model and scalable optimization framework enable a flexible trade-off between performance and complexity, which is vital for realizing the performance gains promised by large IRS-assisted communication systems in practice.
%To this end, we adopt a physics-based IRS model which, unlike the over-idealized conventional IRS model, takes into account the impact of the incident and reflection angles of the impinging electromagnetic waves on the reflected signals.
\end{abstract}
%\vspace*{-1mm}
\section{Introduction}
%\vspace*{3mm}
Next-generation wireless networks are envisioned to offer sustainable high data-rate communication services. To satisfy this demand, radio frequency (RF) transmission-enabled simultaneous wireless information and power transfer (SWIPT) has been proposed as a promising technique for prolonging the lifetime of energy-constrained communication systems \cite{wong2017key}, \cite{6781609}. However, as the signal attenuation associated with the path loss increases with the transmission distance, the received power may not be adequate to ensure stable operation of power-hungry devices, unless they are located very close to the wireless energy transmitter \cite{6781609}. Moreover, due to the random nature of wireless channels, the performance of SWIPT systems can be severely degraded when the radio propagation environment is unfavorable. Indeed, these issues can potentially jeopardize the provision of high data-rate and sustainable communication services creating a performance bottleneck for SWIPT systems.
\par
To overcome these challenges, intelligent reflecting surfaces (IRSs) have been recently advocated for application in SWIPT systems \cite{9136592,pan2020reconfigurable,8910627,9140329}. In particular, exploiting their programmability, the IRS elements can be adjusted to reflect the incident signal with a certain desired phase shift according to the channel conditions \cite{9136592}, \cite{cui2014coding}. As a result, IRSs can be intelligently configured to enhance the received power by constructively combining the signals reflected by different IRS elements at the desired energy harvesting receivers (ERs) or alternatively destructively amalgamating the undesired interference at the information decoding receivers (IRs). This flexibility allows the system designer to customize a favorable radio propagation environment for performance improvement \cite{pan2020reconfigurable}. Moreover, since IRSs comprise passive components with low-power consumption, adding an IRS to the communication infrastructure does not lead to a significant additional energy burden \cite{8910627}, \cite{9140329}. Inspired by these advantages, several works have considered the combination of IRS and SWIPT \cite{9133435,pan2020intelligent,9238963}. For instance, the authors of \cite{9133435} studied the joint design of the beamforming vector at the base station (BS) and the discrete phase shift patterns of the IRS elements for minimization of the BS transmit power in an IRS-aided SWIPT system. Also, in \cite{pan2020intelligent}, the authors considered an IRS-enabled multiple-input multiple-output (MIMO) SWIPT system and developed an alternating optimization (AO)-based algorithm for maximization of the system spectral efficiency while providing reliable wireless power transfer service to multiple ERs. Besides, the authors of \cite{9238963} proposed to jointly optimize the phase shift matrix of the IRS and the beamforming vectors at the BS for security provisioning in an IRS-aided SWIPT system. However, the authors of \cite{9133435} and \cite{pan2020intelligent} adopted an overly-simplified EH model, in which the harvested power of the ERs is linearly proportional to the received RF power. In fact, according to practical field measurements \cite{valenta2014harvesting}, \cite{le2008efficient}, the linear EH model is only accurate when the received RF power is constant. However, due to the combination of the signals from the direct link and the reflected link, which are both fading, the received RF power at the ERs in IRS-assisted SWIPT systems usually has a larger dynamic range than that in conventional SWIPT systems. As a result, the schemes proposed in \cite{9133435} and \cite{pan2020intelligent} may not provide satisfactory wireless power charging service for practical ERs. On the other hand, the authors of \cite{9133435,pan2020intelligent,9238963} adopted an element-wise optimization framework for IRS design such that the computational complexity of the developed optimization algorithms scales with the number of IRS elements. Hence, these algorithms may not be efficient and scalable for online optimization of large IRSs.
\par
Nevertheless, in practice, the number of IRS phase shift elements deployed in future wireless systems is expected to be large. On the one hand, since the phase shifters are usually sub-wavelength elements, a typical rectangular IRS naturally consists of hundreds of elements due to its highly integrated architecture \cite{ma2020information,pei2021prototype,9320594}. For instance, the authors in \cite{pei2021prototype} designed and manufactured a $80$ cm $\times~30$ cm experimental IRS system comprising $1$,$100$ phase shift elements, while an $1$ m $\times~1$ m large IRS prototype composed of $10$,$000$ phase shift elements was demonstrated in \cite{9320594}. On the other hand, even in free space propagation environments, the equivalent path loss of the BS-IRS-receiver link is in general much larger than that of the unobstructed direct link due to the double-path loss effect \cite{di2020analytical}. Hence, to fully realize the potential of IRSs, it is necessary to deploy a large number of phase shift elements such that the severe end-to-end path loss of the cascaded IRS channel can be compensated \cite{wu2020intelligent}, \cite{bjornson2020reconfigurable}. However, with the commonly adopted element-wise optimization framework, both the computational complexity of the existing algorithms, e.g., \cite{pan2020intelligent}, \cite{yu2020irs}, \cite{wu2021intelligent}, and the required signaling overhead grow with the number of IRS elements. As a result, with the element-wise optimization framework, online design of large IRSs may not be feasible in practice, which constitutes a bottleneck for unleashing the full potential of IRSs in wireless communication systems. Therefore, it is necessary to develop an efficient and scalable optimization framework that paves the way to real-time online design for practical IRSs. To address this issue, recently, the authors of \cite{najafi2020physics} developed a physics-based IRS model and a corresponding tile and transmission mode (TT)-based optimization framework. In particular, they proposed to partition the set of IRS elements into several subsets, referred to as tiles, and modeled the impact of each tile on the effective end-to-end wireless channel taking into account the incident angle, the reflection angle, and the polarization of the electromagnetic wave. Subsequently, they developed a scalable optimization framework comprising an offline design stage and an online optimization stage. In the offline stage, the IRS elements of each tile are jointly designed to support a set of
different transmission modes, where each transmission mode effectively corresponds to a given configuration
of the phase shifts. To facilitate efficient online design, the authors of \cite{najafi2020physics} proposed to refine the offline transmission mode set based on a transmission mode pre-selection criterion. Then, in the online stage, the best transmission mode is selected from the refined set according to the design objective. With this new optimization framework, the computational complexity needed for designing large IRSs scales with two design parameters, namely, the number of tiles and the size of the refined transmission mode set. To illustrate this, the authors of \cite{najafi2020physics} considered an IRS-assisted multiple-input single-output (MISO) communication system and developed two efficient algorithms respectively employing AO and greedy approaches for minimization of the BS transmit power subject to quality-of-service (QoS) constraints for IRs. However, these algorithms are not applicable to the IRS-assisted SWIPT systems considered in this paper due to the coexistence of IRs and ERs and the non-linearity of practical EH models. Moreover, the AO-based and greedy algorithms developed in \cite{najafi2020physics} are suboptimal algorithms, while optimal algorithms for the TT-based optimization framework have not been investigated in the literature, yet. Furthermore, the authors of \cite{najafi2020physics} took into account neither user fairness nor specific resource allocation optimization objectives for transmission mode pre-selection. As a result, the schemes proposed in \cite{najafi2020physics} may not fully exploit the benefits of TT-based IRS optimization. Besides, the impact of the tunable parameters, i.e., the number of tiles and the size of the refined transmission mode set, on the computational complexity of the optimization algorithm and the system performance has not be comprehensively investigated in \cite{najafi2020physics}.
\par
In this paper, we address the above issues. The contributions of this paper can be summarized as follows:
\begin{itemize}
\item We investigate the optimal resource allocation algorithm design for large IRS-assisted SWIPT systems based on a realistic non-linear EH model for the ERs and a physics-based IRS model. 
\item We study a TT-based scalable two-stage optimization framework comprising an offline design stage and an online optimization stage. To facilitate efficient online design of IRS-assisted SWIPT systems, we develop two new transmission mode pre-selection criteria to refine the offline transmission mode set.
\item Based on the refined transmission mode set, we jointly optimize the BS beamforming and IRS transmission mode selection policy for minimization of the total transmit power under QoS constraints for both the IRs and the ERs. Although the resulting problem is a non-convex mixed-integer optimization problem, we solve it optimally by exploiting a branch-and-bound (BnB) approach and obtain the optimal online joint beamforming and transmission mode selection strategy. 
\item Since the optimal scheme entails a high computational complexity, we also develop a computationally efficient algorithm by capitalizing on the penalty method, successive convex approximation (SCA), and semidefinite relaxation (SDR). This algorithm asymptotically converges to a locally optimal solution of the considered problem.
\item Simulation results show that the performance of the proposed suboptimal scheme closely approaches that of the proposed optimal scheme. Moreover, our results reveal that the proposed optimal and suboptimal schemes require much lower power consumption compared to three baseline schemes. Furthermore, our results unveil that by properly adjusting the number of tiles and the size of the transmission mode set, we can strike a balance between computational complexity and system performance. Besides, we verify that the proposed scalable optimization framework indeed facilitates an efficient online optimization of large IRS.
\end{itemize}
\par
The remainder of this paper is organized as follows. In Section \ref{Large_IRS_system_model}, we introduce the considered IRS-assisted SWIPT system model. In Section \ref{Large_IRS_optimization_framework}, we first present the adopted TT-based optimization framework, and then develop several transmission mode pre-selection criteria and formulate the resource allocation optimization problem for the considered system. The optimal and suboptimal online joint beamforming and transmission mode selection algorithm designs are provided in Sections \ref{Large_IRS_solution}. In Section \ref{Large_IRS_simulation}, simulation results are presented, and Section \ref{Large_IRS_conclusion} concludes the paper.
\par
\textit{Notation:} In this paper, boldface lower case and boldface capital letters denote vectors and matrices, respectively. $\mathbf{1}_L$ denote the all-ones vector of length $L$. $\mathbb{C}^{N\times M}$ denotes the space of $N\times M$ complex-valued matrices. $\mathbb{H}^{N}$ denotes the set of all $N$-dimensional complex Hermitian matrices. $\mathbf{I}_{N}$ refers to the $N\times N$ identity matrix. $|\cdot|$ and $||\cdot||_2$ denote the absolute value of a complex scalar and the $l_2$-norm of a vector, respectively. $\mathbf{A}^H$ stands for the conjugate transpose of matrix $\mathbf{A}$. $\mathbf{A}\succeq\mathbf{0}$ indicates that $\mathbf{A}$ is a positive semidefinite matrix. $\mathrm{Rank}(\mathbf{A})$ and $\mathrm{Tr}(\mathbf{A})$ denote the rank and the trace of matrix $\mathbf{A}$, respectively. $\mathrm{exp}(x)$ represents the exponential function of a real-valued scalar $x$. $\mathcal{E}\left \{ \cdot \right \}$ denotes statistical expectation. $\sim$ and $\overset{\Delta }{=}$ stand for ``distributed as'' and ``defined as'', respectively. The distribution of a circularly symmetric complex Gaussian random variable with mean $\mu$ and variance $\sigma^2$ is denoted by $\mathcal{CN}(\mu ,\sigma^2)$. $\mathbf{x}^*$ denotes the optimal value of optimization variable $\mathbf{x}$. The gradient vector of function $L(\mathbf{x})$ with respect to $\mathbf{x}$ is denoted by $\nabla_{\mathbf{x}} L(\mathbf{x})$.
%$\left \lceil x \right \rceil$ and $\left \lfloor x \right \rfloor$ denote the rounding up operation and rounding down operation for a real-valued scalar $x$, respectively.
\section{System Model}
\label{Large_IRS_system_model}
%%%%%%%%%%%%%%%%%%%%%%%%%%%%%%%%%%%%%%%%%%%%%%%%%%%%%%%%%%%%%%%%
%\begin{figure}[t]
%\vspace*{0mm}
%\centering
%\includegraphics[width=2.4in]{Large_IRS_SWIPT_System.eps} \vspace*{0mm}
%\caption{An IRS-aided SWIPT system comprising one multi-antenna base station (BS), $K=2$ information decoding receivers (IRs), and $J=2$ energy harvesting receivers (ERs). To facilitate efficient online design, the large IRS is partitioned into $T=8$ tiles of equal size, as indicated by the red-colored dotted boxes.}
%\label{system_model}\vspace*{0mm}
%\end{figure}
%%%%%%%%%%%%%%%%%%%%%%%%%%%%%%%%%%%%%%%%%%%%%%%%%%%%%
\begin{figure}[t]
\centering
\hspace*{-4mm}
\begin{minipage}[b]{0.47\linewidth}
    \centering
\hspace*{-2mm}
\includegraphics[width=2.6in]{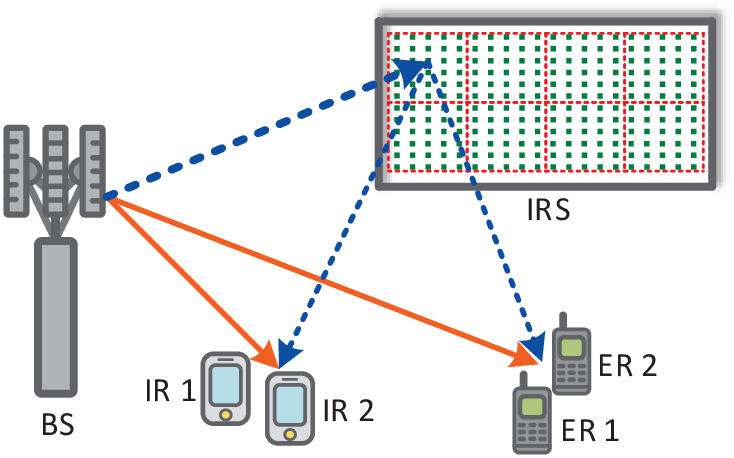}
\vspace*{-6mm}
\caption{An IRS-aided SWIPT system comprising one multi-antenna base station (BS), $K=2$ information decoding receivers (IRs), and $J=2$ energy harvesting receivers (ERs). To facilitate efficient online design, the large IRS is partitioned into $T=8$ tiles of equal size, as indicated by the red-colored dotted boxes.}
\label{system_model}
\end{minipage}
\hspace*{4mm}
\begin{minipage}[b]{0.47\linewidth}
    \centering 
\hspace*{-2mm}
\includegraphics[width=2.4in]{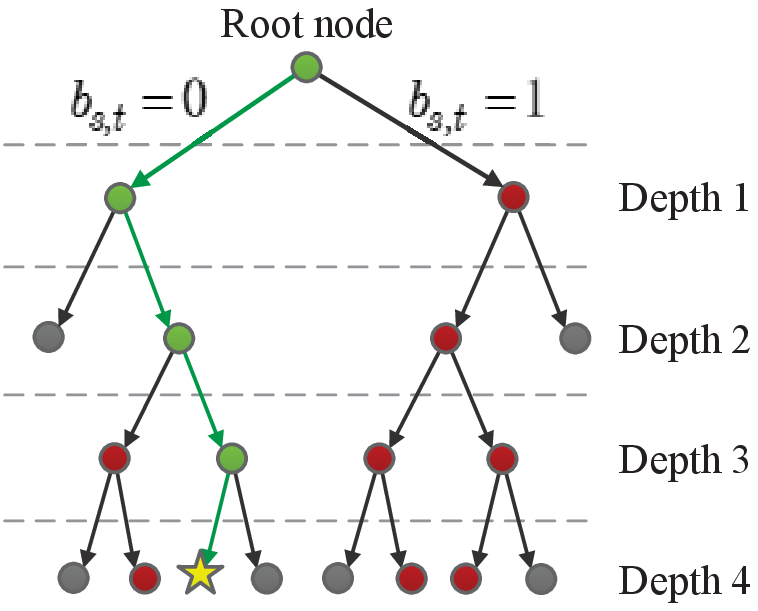}
\vspace*{-6mm}
\caption{An illustration of the BnB search tree for $S=2$ and $T=2$. The green arrows and dots correspond to the path to the optimal node (yellow star). The red and grey dots correspond to non-optimal feasible nodes and discarded nodes, respectively.}\label{BnBfigure}
\end{minipage}
\vspace*{-6mm}
\end{figure}
%%%%%%%%%%%%%%%%%%%%%%%%%%%%%%%%%%%%%%%%%%%%%%%%%%%%%%%%%%%%%%%%%%%%%%%
We consider an IRS-assisted SWIPT system. The system comprises a BS, $K$ IRs, and $J$ ERs, cf. Figure \ref{system_model}. In particular, the BS is equipped with $N_{\mathrm{T}}$ antennas while all receivers are single-antenna devices. To enhance the system performance, a large IRS comprising $M$ phase shift elements is deployed to assist the BS in providing SWIPT services for the two sets of receivers. %In particular, we consider a large IRS of size $M_{x}^{\mathrm{tot}}\times M_{y}^{\mathrm{tot}}$ placed in the $x$-$y$ plane where $M_{x}^{\mathrm{tot}}$, $M_{y}^{\mathrm{tot}}\gg\lambda$. The IRS is composed of many sub-wavelength elements, e.g., phase shifters, of size $M_{e}\times M_{e}$ and each element can reflect the incident signal with a desired phase shift. We assume that the IRS is partitioned into tiles of size $M_{x}\times M_{y}$. For notational simplicity, we assume that the distance between two adjacent elements is negligible and $M_{x}^{\mathrm{tot}}/M_{x}$, $M_{y}^{\mathrm{tot}}/M_{x}$, $M_{x}/M_{e}$, and $M_{y}/M_{e}$ are integers. As a result, there are $T=M_{x}^{\mathrm{tot}}\times M_{y}^{\mathrm{tot}}/(M_{x}\times M_{y})$ tiles and each tile is composed of several programmable elements.
For notational simplicity, we define sets $\mathcal{J}=\left \{1,\cdots ,J \right \}$ and $\mathcal{K}=\left \{1,\cdots ,K \right \}$ to collect the indices of the ERs and IRs, respectively.
\subsection{Signal Model}
In a given scheduling time slot, the BS transmit signal is given by
\begin{equation}
    \mathbf{x}=\underset{k\in\mathcal{K} }{\sum }\mathbf{w}_kd_k+\mathbf{v},
\end{equation}
where $\mathbf{w}_k\in \mathbb{C}^{\mathit{N}_{\mathrm{T}}\times 1}$ and $d_k\in \mathbb{C}$ denote the beamforming vector for IR $k$ and the corresponding information symbol, respectively. Without loss of generality, we assume $\mathcal{E}\{\left |d_k \right|^2\}=1$, $\forall\mathit{k} \in \mathcal{K}$. Moreover, $\mathbf{v}$ is generated as a Gaussian pseudo-random sequence which is utilized to provide wireless power transfer service for ERs and is known to both the IRs and ERs \cite{6781609}, \cite{6860253}. In particular, $\mathbf{v}$ is modeled as 
\begin{equation}
    \mathbf{v}\sim \mathcal{CN}\left ( \mathbf{0}, \mathbf{V} \right ),
\end{equation}
where $\mathbf{V}\succeq\mathbf{0}$, $\mathbf{V}\in\mathbb{H}^{N_{\mathrm{T}}}$, denotes the covariance matrix of the pseudo-random energy signal.
\par
%\begin{Remark}
%We note that unlike other works in \cite{9148892} and \cite{pan2020intelligent} that assign one beamformer to each ER, we adopt an energy signal $\mathbf{v}$ which is able to simultaneously charge all ERs. 
%\end{Remark}
In each scheduling time slot, the received signals at IR $k$ and ER $j$ are given by
\begin{eqnarray}
    y_{\mathrm{I},k}&\hspace*{-3mm}=\hspace*{-3mm}&\underbrace{\mathbf{h}^H_{\mathrm{I},k}\mathbf{w}_kd_k}_{\text{desired signal}}+\underbrace{\mathbf{h}^H_{\mathrm{I},k}\underset{r\in\mathcal{K}\setminus \left \{ k \right \} }{\sum }\mathbf{w}_rd_r}_{\text{multiuser interference}}+\underbrace{\mathbf{h}^H_{\mathrm{I},k}\mathbf{v}}_{\text{energy signal}}+n_{\mathrm{I},k},\\
    y_{\mathrm{E},j}&\hspace*{-3mm}=\hspace*{-3mm}&\mathbf{h}^H_{\mathrm{E},j}\left ( \underset{k\in\mathcal{K} }{\sum }\mathbf{w}_kd_k+\mathbf{v}\right )+n_{\mathrm{E},j},
\end{eqnarray}
respectively. Here, variables $n_{\mathrm{I},k}\sim\mathcal{CN}(0,\sigma_{\mathrm{I}_k}^2)$ and $n_{\mathrm{E},j}\sim\mathcal{CN}(0,\sigma_{\mathrm{E}_j}^2)$ denote the additive white Gaussian noise with variance $\sigma_{\mathrm{I}_k}^2$ and $\sigma_{\mathrm{E}_j}^2$ at IR $k$ and ER $j$, respectively. Moreover, vectors $\mathbf{h}_{\mathrm{I},k}$ and $\mathbf{h}_{\mathrm{E},j}$ respectively denote the effective BS-IR$_k$ and BS-ER$_j$ channels
which are the superpositions of the channels of direct links and IRS-assisted links, i.e., 
\begin{equation}
	\mathbf{h}_{\mathrm{I},k}=\mathbf{h}_{\mathrm{I},k,\mathrm{R}}+\mathbf{h}_{\mathrm{I},k,\mathrm{D}}~~\mbox{and}~~
	\mathbf{h}_{\mathrm{E},j}=\mathbf{h}_{\mathrm{E},j,\mathrm{R}}+\mathbf{h}_{\mathrm{E},j,\mathrm{D}}.
\end{equation}
Here, $\mathbf{h}_{\mathrm{I},k,\mathrm{R}}$ and $\mathbf{h}_{\mathrm{E},j,\mathrm{R}}$ are the IRS-assisted BS-IR$_k$ and BS-ER$_j$ channels, respectively, and $\mathbf{h}_{\mathrm{I},k,\mathrm{D}}$ and $\mathbf{h}_{\mathrm{E},j,\mathrm{D}}$ denote the channels of the corresponding direct links. Further details regarding the adopted channel models are presented in Section III-A.   
Besides, to investigate the maximum achievable performance, similar to \cite{8930608}, \cite{9076830}, we assume that the perfect channel state information (CSI) of the entire system is available at the BS.
%\footnote{An efficient channel estimation framework for the adopted physics-based IRS model has been proposed in Section III-D in \cite{najafi2020physics}.}
\par
%For instance, to design the configuration of the phase shifts of an IRS with $1000$ elements, the computational complexity of the resource allocation algorithm for the conventional IRS model is proportional to the $1000$ elements which may not be feasible for online optimization. While by adopting the physics-based model and scalable optimization of the large IRS, we partition the IRS into $5$ tiles of equal size and jointly design the $200$ phase shift elements in each tile for the support of $10$ transmission modes, where each transmission mode effectively manipulate the incident signals.
\subsection{Non-linear Energy Harvesting Model}
Most existing works on IRS-assisted SWIPT systems adopted a linear EH model \cite{pan2020intelligent}, \cite{9148892}. However, this model is not accurate as the RF energy conversion efficiency depends on the input RF power level of the EH circuit. To capture this effect, in this paper, we adopt the non-linear EH model proposed in \cite{boshkovska2015practical}. In particular, the energy harvested by ER $j$, i.e., $\Upsilon_j^{\mathrm{EH}}$, is modeled by
\begin{eqnarray}
    \Upsilon_j^{\mathrm{EH}}&\hspace*{-2mm}=\hspace*{-2mm}&\frac{\Lambda_j-a_j\Xi_j}{1-\Xi_j}, ~~\Xi_j=\frac{1}{1+\mathrm{exp}(\varrho _j c_j)},\\
    \Lambda_j&\hspace*{-2mm}=\hspace*{-2mm}&\frac{a_j}{1+\mathrm{exp}\left(-\varrho _j\left ( P^{\mathrm{ER}}_j-c_j \right )\right)},
\end{eqnarray}
where $\Lambda_j$ is a logistic function whose input is the received RF power $P^{\mathrm{ER}}_j$ at ER $j$, and constant $\Xi_j$ ensures a zero-input/zero-output response. Besides, $a_j$, $c_j$, and $\varrho _j$ are constant parameters that depend on the employed EH circuit. Given the schematic of the EH circuit, these parameters can be easily determined via some standard curve fitting method \cite{boshkovska2015practical}.
%%%%%%%%%%%%%%%%%%%%%%%%%%%%%%%%%%%%%%%%%%%%%%%%%%
\section{A Scalable Optimization Framework for IRS-assisted SWIPT Systems}
\label{Large_IRS_optimization_framework}
In this section, we first present the adopted IRS model which enables a scalable optimization of large IRSs. To facilitate an efficient online design for larger IRSs, we also develop three transmission mode pre-selection criteria for large IRS-assisted SWIPT systems. At last, after introducing the adopted performance metrics, we formulate the resource allocation design as an optimization problem.
\subsection{End-to-End IRS Channel Model Based on TT Framework}
\label{IRS model}
In the following, we adopt the TT-based optimization framework recently developed in \cite{najafi2020physics} to the considered IRS-assisted SWIPT system. We first briefly present the general form of the TT-based framework and then specialize it to the case where a physics-based model is used to characterize the wireless channel and the IRS. Finally, we discuss why this framework enables scalable optimization.

\subsubsection{TT Framework} In practice, optimizing the individual phase shift elements may not be affordable for online design of large IRS. To address this issue, in the TT framework, \textit{i)} the large IRS is partitioned into $T$ tiles of equal sizes and \textit{ii)} a set of $S$ phase shift configurations for all IRS elements of each tile is designed offline \cite{najafi2020physics}, \cite{yu2021smart}. Throughout this paper, we refer to the phase shift configurations as transmission modes and employ the same set of transmission modes for all IRS tiles. Let $\mathbf{h}_{\mathrm{I},k,s,t,\mathrm{R}}$ and $\mathbf{h}_{\mathrm{E},j,s,t,\mathrm{R}}$ denote the end-to-end  BS-IR$_k$ and BS-ER$_j$ channels, respectively, of tile $t$ and realized by transmission mode $s$. Then, $\mathbf{h}_{\mathrm{I},k,\mathrm{R}}$ and $\mathbf{h}_{\mathrm{E},j,\mathrm{R}}$ are given by, respectively,
\begin{equation}\label{eq:TTmodel}
	\mathbf{h}_{\mathrm{I},k,\mathrm{R}}=\underset{\substack{s\in\mathcal{S},t\in\mathcal{T}}}{\sum }b_{s,t}\mathbf{h}_{\mathrm{I},k,s,t,\mathrm{R}}~~\mbox{and}~~
	\mathbf{h}_{\mathrm{E},j,\mathrm{R}}=\underset{\substack{s\in\mathcal{S},t\in\mathcal{T}}}{\sum }b_{s,t}\mathbf{h}_{\mathrm{E},j,s,t,\mathrm{R}},
\end{equation}
where sets $\mathcal{S}=\left \{ 1,\cdots ,S \right \}$ and $\mathcal{T}=\left \{ 1,\cdots ,T \right \}$ collect the indices of the transmission modes and tiles, respectively. Furthermore, $b_{s,t}\in\left\{0,1\right\}$ is a binary variable with $b_{s,t}=1$ if tile $t$ employs transmission mode $s$, $\forall s\in\mathcal{S}$, $\forall t\in\mathcal{T}$; otherwise it is equal to zero. Constraint $\underset{\substack{s\in\mathcal{S}}}{\sum }b_{s,t}=1$, $\forall t$, has to hold since only one transmission mode can be selected for each tile. Note that for a given set of transmission modes, channel vectors $\mathbf{h}_{\mathrm{I},k,s,t,\mathrm{R}}$ and $\mathbf{h}_{\mathrm{E},j,s,t,\mathrm{R}}$ are fixed. However, unlike conventional systems, we can choose among $S^T$ possible end-to-end IRS-assisted channels by optimizing $b_{s,t}$. 

For ease of presentation and without loss of generality, we model the direct link via a virtual tile indexed by $t=0$ and rewrite $\mathbf{h}_{\mathrm{I},k,\mathrm{D}}$ as $\mathbf{h}_{\mathrm{I},k,\mathrm{D}}=\underset{\substack{s\in\mathcal{S}}}{\sum }b_{s,0}\mathbf{h}_{\mathrm{I},k,s,0}$ with $\mathbf{h}_{\mathrm{I},k,s,0}=\mathbf{h}_{\mathrm{I},k,\mathrm{D}},\,\,\forall s$. 
By doing so, we can simplify the notation by dropping the subscripts $\mathrm{R}$ and $\mathrm{D}$ in all the channel vectors. We note that for ER $j$,  channel vector  $\mathbf{h}_{\mathrm{E},j,\mathrm{D}}$ can be defined in a similar manner as  $\mathbf{h}_{\mathrm{I},j,\mathrm{D}}$. As a result, the effective end-to-end BS-IR$_k$ and BS-ER$_j$ channels can be rewritten as, respectively,
\begin{equation}
	\mathbf{h}_{\mathrm{I},k}=\underset{\substack{s\in\mathcal{S},t\in\mathcal{\widehat{\mathcal{T}}}}}{\sum }b_{s,t}\mathbf{h}_{\mathrm{I},k,s,t}~~\mbox{and}~~\mathbf{h}_{\mathrm{E},j}=\underset{\substack{s\in\mathcal{S},t\in\mathcal{\widehat{\mathcal{T}}}}}{\sum }b_{s,t}\mathbf{h}_{\mathrm{E},j,s,t},\label{channel1}
\end{equation}
where $\widehat{\mathcal{T}}=\mathcal{T}\cup \left \{ 0 \right \}$.

\subsubsection{Physics-based Model} We assume that the wireless environment consists of multiple scatterers, where each scatterer contributes a single propagation path \cite{najafi2020physics}, \cite{6847111}. Hence, the signal transmitted by the BS will arrive at the IRS via multiple paths, and the signal reflected by the IRS will arrive at a given receiver via multiple paths. To fully unleash the potential of the IRS, it is crucial to reflect the signal along the strong paths in order to ensure sufficient link budget especially when the direct links suffer from severe shadowing. Motivated by this discussion, instead of modeling the IRS tiles in terms of the reflection coefficients of all individual reflecting elements, we model the tile for each transmission mode in terms of its response for all angle of arrivals (AoAs)  corresponding to the paths in the BS-IRS link and angle-of-departures (AoDs) corresponding to the paths in the IRS-receiver link. In particular, for tile $t$ employing transmission mode $s$, the effective end-to-end channel between the BS and IR $k$, i.e., $\mathbf{h}_{\mathrm{I},k,s,t}$, is given by \cite{najafi2020physics}
\begin{equation}
	\mathbf{h}_{\mathrm{I},k,s,t}^H=\mathbf{1}_{\mathrm{I},k,\mathrm{R}}^H\mathbf{C}_{\mathrm{I},k,\mathrm{R}}\mathbf{R}_{\mathrm{I},k,s,t}\mathbf{C}_{\mathrm{T}}\mathbf{D}_{\mathrm{T}},~~\forall t\in\mathcal{T}.\label{channel2}
\end{equation}
Here, $\mathbf{1}_{\mathrm{I},k,\mathrm{R}}\in\mathbb{C}^{L_{\mathrm{I},k,\mathrm{R}}\times 1}$ is an all-ones vector, and $\mathbf{D}_{\mathrm{T}}\in\mathbb{C}^{L_{\mathrm{T}}\times N_{\mathrm{T}}}$ is a matrix representing the BS antenna array response of the BS-IRS link. $L_{\mathrm{T}}$ and $L_{\mathrm{I},k,\mathrm{R}}$ denote the numbers of scatterers of the BS-IRS and IRS-IR$_k$ links, respectively. Moreover, diagonal matrices $\mathbf{C}_{\mathrm{T}}\in\mathbb{C}^{L_{\mathrm{T}}\times L_{\mathrm{T}}}$ and $\mathbf{C}_{\mathrm{I},k,\mathrm{R}}\in\mathbb{C}^{L_{\mathrm{I},k,\mathrm{R}}\times L_{\mathrm{I},k,\mathrm{R}}}$ contain the channel coefficients which capture the joint impact of path loss, shadowing, and small-scale fading on the BS-IRS and IRS-IR$_k$ links, respectively. Furthermore, matrix $\mathbf{R}_{\mathrm{I},k,s,t}\in\mathbb{C}^{L_{\mathrm{I},k,\mathrm{R}}\times L_{\mathrm{T}}}$ denotes the response function of tile $t$ applying the $s$-th transmission mode evaluated at the AoAs of the BS-IRS link and AoDs of the IRS-IR$_k$ link. We note that for each channel realization, $\mathbf{R}_{\mathrm{I},k,s,t}$ is fixed for a given transmission mode and depends on the channel AoAs and AoDs, see \cite[Eq. (26)]{najafi2020physics} and \cite[Eq. (5)]{jamali2020power}. In other words, different transmission modes realize different $\mathbf{R}_{\mathrm{I},k,s,t}$ and consequently different end-to-end channels $\mathbf{h}_{\mathrm{I},k,s,t}$.
\begin{Remark}
We note that the physics-based model in \eqref{channel2} is a more general and more accurate model compared to the widely adopted conventional IRS model \cite{8930608,hu2020sum,9139273}. On the one hand, by setting $T$ equal to the number of IRS elements $M$ and mapping the entire phase shift domain to the transmission mode set, the conventional IRS model can be regarded as a special case of the IRS model considered in this paper. On the other hand, unlike the conventional IRS model over-optimistically assuming a constant gain for all signals reflected by the IRS, the physics-based model takes into account the impact of the incident and reflection angles of the impinging electromagnetic waves on the reflected signals when determining the corresponding gain.
\end{Remark}
\par
Assuming scatterer-based wireless environment, the channel vector of the direct link between the BS and IR$_k$ is given by \cite{najafi2020physics}
\begin{equation}
	\mathbf{h}_{\mathrm{I},k,0}^H=\mathbf{1}_{\mathrm{I},k,\mathrm{D}}^H\mathbf{C}_{\mathrm{I},k,\mathrm{D}}\mathbf{D}_{\mathrm{I},k,\mathrm{D}},\label{eq:direct}
\end{equation}
where $\mathbf{1}_{\mathrm{I},k,\mathrm{D}}\in\mathbb{C}^{L_{\mathrm{I},k,\mathrm{D}}\times 1}$ is an all-ones vector, $\mathbf{D}_{\mathrm{I},k,\mathrm{D}}\in\mathbb{C}^{L_{\mathrm{I},k,\mathrm{D}}\times N_{\mathrm{T}}}$ is a matrix containing the BS antenna array vectors of the BS-IR$_k$ direct link,  $\mathbf{C}_{\mathrm{I},k,\mathrm{D}}\in\mathbb{C}^{L_{\mathrm{I},k,\mathrm{D}}\times L_{\mathrm{I},k,\mathrm{D}}}$ is a diagonal matrix whose diagonal entries are the channel coefficients of the scatterers in the BS-IR$_k$ direct link, and $L_{\mathrm{I},k,\mathrm{D}}$ denotes the number of scatterers of the BS-IR$_k$ direct link. The BS-ER$_j$ channel realized via tile $t$ for transmission mode $s$, i.e., $\mathbf{h}_{\mathrm{E},j,s,t}$, and the BS-ER$_j$ direct channel $\mathbf{h}_{\mathrm{E},j,0}$ are modeled similar to $\mathbf{h}_{\mathrm{I},k,s,t}$ and $\mathbf{h}_{\mathrm{I},k,0}$ in \eqref{channel2} and \eqref{eq:direct}, respectively. 

\subsubsection{Two-Stage Scalable Optimization Framework} 
IRS optimization based on \eqref{eq:TTmodel} and \eqref{channel2} is performed in two stages, namely an offline stage and an online stage. \textit{i)} In the offline stage, the set of transmission modes is designed which determines the set of potential end-to-end channels that each tile can create for a given channel realization. \textit{ii)} In the online stage, the best transmission mode is chosen for each tile depending on the design objective of the considered systems. The scalability of this TT framework stems from the fact that the computational complexity for online IRS optimization is untied from the number of IRS elements $M$, but scales with the number of tiles $T$ and the number of transmission modes $S$ which are design parameters and can be adjustable to trade performance for complexity. The problem of offline transmission mode set design has been previously studied in \cite{najafi2020physics} and \cite{jamali2020power}, and the proposed transmission mode sets are also applicable to the IRS-assisted SWIPT systems considered here. Therefore, in this paper, we focus our attention on the online optimization stage. Nevertheless, we emphasize that the online optimization algorithms developed in this paper are applicable for any offline transmission mode set.

\subsection{Transmission Mode Pre-selection for Online Optimization}\label{Pre-selection}
In this paper, we assume that the transmission mode set is generated offline according to \cite[Section III-A]{najafi2020physics} and is then employed for online optimization of the considered system. However, before performing online resource allocation, it is advisable to refine the obtained transmission mode set to facilitate a more computationally efficient online optimization. Therefore, in the following, we first present a simple transmission mode pre-selection criterion which was proposed in \cite{najafi2020physics}. Noticing the drawback of this criterion, we develop two new criteria taking into account user fairness and the specific characteristics of SWIPT systems, respectively.
\par
For the effective end-to-end channel given in \eqref{channel2}, a given channel realization does not only depend on the channel coefficients of the links but also on the locations of the scatterers. In fact, since the number of scatterers is limited and the locations of the scatterers are fixed, the IRS can exploit only a limited number of AoAs and AoDs to receive the signals from the BS or to reflect the signals to the receivers. As a result, only a limited number of transmission modes are suitable candidates for reflecting the signal impinging from one of the scatterers/BS towards one of the other scatterers/receivers \cite{najafi2020physics}, \cite{jamali2020power}. Thus, for a given BS-receiver pair, the Euclidean norm of the channel vector, i.e., $\left \| \mathbf{h}_{\mathrm{I},k,s,t} \right \|_2$ or $\left \| \mathbf{h}_{\mathrm{E},j,s,t} \right \|_2$, is non-negligible only for a few of the transmission modes contained in the offline transmission mode set. Thus, to facilitate online optimization for practically large IRS-assisted SWIPT systems, we propose to first pre-select a subset of the transmission modes from the set generated in the offline design stage. 
\par
The authors of \cite{najafi2020physics} developed a simple and straightforward transmission mode pre-selection criterion which chooses the transmission modes that yield the largest effective end-to-end channel gain, i.e., the largest Euclidean norm of the channel vector. In particular, the desired refined transmission mode set\footnote{For notational simplicity, in the rest of the paper, we slightly abuse the notation and denote the refined transmission mode set and the number of transmission modes after pre-selection also by $\mathcal{S}$ and $S$, respectively.} can be obtained as follows
%\footnote{For easy of presentation, in this subsection, we slightly abuse the notation of the channel vector $\mathbf{h}_{s,t,i}$ to also denote either $\mathbf{h}_{\mathrm{I},k,s,t}$ or $\mathbf{g}_{s,t,j}$.}
\begin{equation}
    \mbox{Criterion~1:~}\mathcal{S}=\left \{s|\exists (r,i,t):\left \| \mathbf{h}_{r,i,s,t} \right \|_2\geq \delta_1, \left\{\begin{matrix}
r=\mathrm{I}, & i\in\mathcal{K} \\ 
r=\mathrm{E}, & i\in\mathcal{J}
\end{matrix}\right\},\forall t\in\mathcal{T} \right \},
\end{equation}
where $\delta_1>0$ is a tunable threshold which can be used to adjust the size of set $\mathcal{S}$. Although Criterion 1 is simple to implement, it does not take into account user fairness and may lead to a biased refined transmission mode set that is favorable only for one IR or one ER. In fact, if the channel state of one receiver is much better than that of the other receivers, it is possible that most of the phase-shift configurations in the refined transmission mode set are only favorable for this receiver. To circumvent this issue, we can construct an improved transmission mode set by pre-selecting a few favorable phase-shift configurations for each receiver. Specifically, the transmission mode set for a given receiver is constructed as follows
\begin{equation}
    \mbox{Criterion~2:~}\mathcal{S}_{r,i}=\left \{s~|~\exists t:\left \| \mathbf{h}_{r,i,s,t} \right \|_2\geq \delta_{2,r,i},\forall t\in\mathcal{T} \right \}, \left\{\begin{matrix}
r=\mathrm{I}, & i\in\mathcal{K} \\ 
r=\mathrm{E}, & i\in\mathcal{J}
\end{matrix}\right., 
\end{equation}
where $\delta_{2,r,i}>0$ is a parameter to adjust the size of set $\mathcal{S}_{r,i}$ and the total set in this case is given by $\mathcal{S}=\underset{r,i }{\bigcup}\mathcal{S}_{r,i}$.
\par
Apart from the aforementioned criteria, one can also pre-select transmission modes based on the specific resource allocation optimization objective. Considering the fact that in SWIPT systems, the ERs usually require much higher received powers than the IRs, we can refine the transmission mode set by keeping those transmission modes that are favorable for the ERs, which facilitates minimization of the BS transmit power while providing satisfactory service for both IRs and ERs. This can be achieved by imposing weights on the effective end-to-end channels of the ERs. To this end, the transmission mode set for receiver $i$ is constructed based on the following weight criterion
\begin{equation}
\mbox{Criterion~3:~}\mathcal{S}=\left \{s~|~\exists t:p_{r,i,s,t}\geq \delta_1,r\in\left \{ \mathrm{I}, \mathrm{E} \right \}, \forall t\in\mathcal{T}\right \},
\end{equation}
where variable $p_{r,i,s,t}$ is defined as
\begin{equation}
\label{weightedchannel}
p_{r,i,s,t}=\left\{\begin{matrix}
\omega\left \|  \mathbf{h}_{r,i,s,t} \right \|_2, &r=\mathrm{E},~\forall i\in\mathcal{J} \\ 
\left \|\mathbf{h}_{r,i,s,t} \right \|_2, &r=\mathrm{I},~\forall i\in\mathcal{K}
\end{matrix}\right..
\end{equation}
Here, $\omega\geq 1$ is the weight factor to be used to prioritize the ERs such that the resulting set is more favorable for ERs. Given the refined transmission mode set, we can proceed to formulate the online optimization problem in the next subsection.
\subsection{Optimization Problem Formulation}
%\begin{figure*}[t]
%\setcounter{TempEqCnt}{\value{equation}} 		 
	%将当前公式序号 赋给TempEqCnt
	%\setcounter{equation}{\value{equation}}
%\begin{eqnarray}
%\label{SINR}
    %\hspace*{-6mm}\Gamma_k&\hspace*{-3mm}=\hspace*{-3mm}&\frac{\underset{\substack{s\in\mathcal{S},t\in\widehat{\mathcal{T}}}}{\sum }\underset{\substack{p\in\mathcal{S},q\in\widehat{\mathcal{T}}}}{\sum }b_{s,t}b_{p,q}\mathbf{h}^H_{s,t,k} \mathbf{w}_k\mathbf{w}^H_k\mathbf{h}_{p,q,k}}{\underset{\substack{s\in\mathcal{S},t\in\widehat{\mathcal{T}}}}{\sum }\underset{\substack{p\in\mathcal{S},q\in\widehat{\mathcal{T}}}}{\sum }b_{s,t}b_{p,q}\mathbf{h}^H_{s,t,k}\left (  \underset{r\in\mathcal{K}\setminus \left \{ k \right \} }{\sum }\mathbf{w}_r\mathbf{w}^H_r+\mathbf{V} \right )\mathbf{h}_{p,q,k}+\hspace*{-0.5mm}\sigma ^2_{\mathrm{I}_k}}.\\
    %\hspace*{-6mm}P^{\mathrm{ER}}_j&\hspace*{-3mm}=\hspace*{-3mm}&\underset{\substack{s\in\mathcal{S},t\in\widehat{\mathcal{T}}}}{\sum }\underset{\substack{p\in\mathcal{S},q\in\widehat{\mathcal{T}}}}{\sum }b_{s,t}b_{p,q}\mathbf{g}^H_{s,t,j}\left (  \underset{k\in\mathcal{K}}{\sum }\mathbf{w}_k\mathbf{w}^H_k+\mathbf{V}\right )\mathbf{g}_{p,q,j}.\label{HarvestedPower}
%\end{eqnarray}
%\hrule
%\end{figure*}
%The receive signal-to-noise-plus-interference ratio (SINR) of IR $k$, i.e., $\Gamma_k$ is given in \eqref{SINR} shown at the top of the page.
The received signal-to-noise-plus-interference ratio (SINR) of IR $k$, i.e., $\Gamma_k$ is given by\footnote{Since the pseudo-random energy signal is perfectly known at all receivers in principle, the corresponding interference can be canceled at the IRs. In this case, the energy signal-induced interference, i.e., $\mathbf{h}_{\mathrm{I},k}^H\mathbf{V}\mathbf{h}_{\mathrm{I},k}$, is absent in the expression for the SINR. On the other hand, applying interference cancellation increases receiver complexity, of course. Without loss of generality, in this paper, we assume that the IRs do not implement cancellation of the energy signal-induced interference. Nevertheless, we note that the schemes proposed in this paper are also applicable to the case where cancellation of the energy signal-induced interference is possible.}
%\begin{eqnarray}
%\label{SINR}
%\Gamma_k&\hspace*{-3mm}=\hspace*{-3mm}&\frac{\underset{\substack{s\in\mathcal{S},t\in\widehat{\mathcal{T}}}}{\sum }\underset{\substack{p\in\mathcal{S},q\in\widehat{\mathcal{T}}}}{\sum }b_{s,t}b_{p,q}\mathbf{h}^H_{s,t,k} \mathbf{w}_k\mathbf{w}^H_k\mathbf{h}_{p,q,k}}{\underset{\substack{s\in\mathcal{S},t\in\widehat{\mathcal{T}}}}{\sum }\underset{\substack{p\in\mathcal{S},q\in\widehat{\mathcal{T}}}}{\sum }b_{s,t}b_{p,q}\mathbf{h}^H_{s,t,k}\left (  \underset{r\in\mathcal{K}\setminus \left \{ k \right \} }{\sum }\mathbf{w}_r\mathbf{w}^H_r+\mathbf{V} \right )\mathbf{h}_{p,q,k}+\hspace*{-0.5mm}\sigma ^2_{\mathrm{I}_k}}.
%\end{eqnarray}
\begin{eqnarray}
\label{SINR}
\Gamma_k&=&\frac{\left |\mathbf{h}_{\mathrm{I},k}^H \mathbf{w}_k\right |^2}{\mathbf{h}_{\mathrm{I},k}^H \big(\underset{\substack{r\in\mathcal{K}\setminus \left \{ k \right \}}}{\sum }\mathbf{w}_r\mathbf{w}^H_r+\mathbf{V}\big)\mathbf{h}_{\mathrm{I},k}+\sigma ^2_{\mathrm{I}_k}}.
\end{eqnarray}
Furthermore, the received RF power at ER $j$, i.e., $P_j^{\mathrm{ER}}$, is given by
%\begin{eqnarray}
%P^{\mathrm{ER}}_j&\hspace*{-3mm}=\hspace*{-3mm}&\underset{\substack{s\in\mathcal{S},t\in\widehat{\mathcal{T}}}}{\sum }\underset{\substack{p\in\mathcal{S},q\in\widehat{\mathcal{T}}}}{\sum }b_{s,t}b_{p,q}\mathbf{g}^H_{s,t,j}\left (  \underset{k\in\mathcal{K}}{\sum }\mathbf{w}_k\mathbf{w}^H_k+\mathbf{V}\right )\mathbf{g}_{p,q,j}.\label{HarvestedPower}
%\end{eqnarray}
%\par
\begin{eqnarray}
P^{\mathrm{ER}}_j&=&\mathbf{h}_{\mathrm{E},j}^H\left (  \underset{k\in\mathcal{K}}{\sum }\mathbf{w}_k\mathbf{w}^H_k+\mathbf{V}\right )\mathbf{h}_{\mathrm{E},j}.\label{HarvestedPower}
\end{eqnarray}
\par
In this paper, we aim to minimize the total transmit power at the BS while satisfying the QoS requirements of the IRs and the EH requirements of the ERs. In particular, the joint beamforming and transmission mode selection policy, i.e., $\left \{\mathbf{w}_k,\mathbf{V},b_{s,t} \right \}$, is obtained by solving the following optimization problem
\begin{eqnarray}
\label{prob1}
&&\hspace*{2mm}\underset{\mathbf{V}\in\mathbb{H}^{\mathit{N}_{\mathrm{T}}},\mathbf{w}_k,b_{s,t}}{\mino} \,\, \,\, \hspace*{2mm}\underset{k\in\mathcal{K}}{\sum }\left \| \mathbf{w}_k \right \|^2+\mathrm{Tr}(\mathbf{V})\notag\\
\mbox{s.t.}\hspace*{-4mm}
&&\mbox{C1:~}\Gamma_k\geq \Gamma_{\mathrm{req}_k},~\forall k,\hspace*{4mm}\mbox{C2:~}\Upsilon_j^{\mathrm{EH}}\geq E_{\mathrm{req}_j},~\forall j,\hspace*{4mm}\mbox{C3:~}\mathbf{V}\succeq\mathbf{0},\notag\\
&&\mbox{C4:~}b_{s,t}\in\left \{ 0,1 \right \},~\forall s,~\forall t,\hspace*{4mm}
\mbox{C5:~}\underset{ s\in\mathcal{S}}{\sum}b_{s,t}=1,~\forall t.
\end{eqnarray}
Here, $\Gamma_{\mathrm{req}_k}$ in constraint C1 is the pre-defined minimum required SINR of IR $k$. Constraint C2 indicates that the minimum harvested power at ER $j$ should be greater than a given threshold $E_{\mathrm{req}_j}$. Constraint C3 and $\mathbf{V}\in\mathbb{H}^{\mathit{N}_{\mathrm{T}}}$ restrict matrix $\mathbf{V}$ to be a positive semidefinite Hermitian matrix such that it is a valid covariance matrix. Constraints C4 and C5 are imposed since only one transmission mode can be selected for each tile.
%\begin{Remark}
%We note that the proposed scalable optimization framework facilitates an efficient design of large IRSs compared to the conventional element-wise optimization framework. Specifically, assuming the effective end-to-end channels in \eqref{channel2} are known, we aim to select the best transmission mode for each tile from the refined transmission mode set, instead of individually optimizing the phase shift of each IRS element. As a result, the computational complexity for IRS design are untied with the number of IRS elements $M$, but scales with with the numbers of tiles and the size of the refined transmission mode set, i.e., $T$ and $S$, which are design parameters and can be adjustable to trade performance for complexity. 
%\end{Remark}
\par
The problem in \eqref{prob1} is a mixed-integer non-convex optimization problem. Although \eqref{prob1} is still challenging to solve due to the non-convexity stemming from the coupling of the optimization variables, the fractional function in constraint C1, and the binary selection constraint in C4, we sidestep the unit-modulus constraint of the optimization problems formulated in \cite{9133435,pan2020intelligent,9238963} based on the conventional element-wise optimization framework. This allows us to leverage the plethora of algorithms developed for integer programming problems to preserve joint optimality rather than solely relying on AO-based algorithms. Therefore, in the next section, we develop a BnB-based algorithm to obtain the globally optimal solution of problem \eqref{prob1} which serves as a performance benchmark. Since the BnB-based algorithm entails a high computational complexity, we also develop a suboptimal scheme which has only polynomial time computational complexity.
%where constraint C1 is a power constraint for the UAV with maximum transmit power allowance $P_{\mathrm{max}}$. In constraints C2, we take into account the quality-of-service and set $\Gamma_{\mathrm{req}_k}$ to be the minimum SINR of user $k$ required for information decoding. $\mathrm{r}_{\mathrm{req}}$ in constraint C3 denotes the minimum distance between the UAV and the information source that is able to maintain continuous and error-free information collection, and $\mathbf{r}_s$ indicates the 3-D Cartesian coordinates of the information source. Constraint C4 is a height requirement of the UAV position to guarantee LoS channels between the UAV and users. The problem in \eqref{OP1} is a non-convex problem due to constraint C2 which involves infinitely many inequality constraints.
%\vspace{-1mm}
\section{Solution of the Problem}
\label{Large_IRS_solution}
In this section, we first tackle the coupling of the optimization variables by employing the big-M formulation \cite{griva2009linear}. Then, a BnB-based algorithm is proposed to solve the optimization problem in \eqref{prob1} optimally leading to an iterative resource allocation algorithm. In each iteration, a non-convex optimization problem is solved optimally by semidefinite relaxation (SDR). Subsequently, we develop a suboptimal solution based on SCA which asymptotically converges to a locally optimal solution of the considered optimization problem in polynomial time.
\subsection{Problem Transformation}
\label{problem transformation}
To facilitate resource allocation algorithm design, we first handle the coupling of the optimization variables by defining $\mathbf{W}_k=\mathbf{w}_k\mathbf{w}_k^H$, $\forall k$. Considering the channel vectors $\mathbf{h}_{\mathrm{I},k}$ and $\mathbf{h}_{\mathrm{E},j}$ defined in \eqref{channel1}, we note that \eqref{SINR} and \eqref{HarvestedPower} contain cross-terms $b_{s,t}b_{p,q}$, $\forall s,p\in\mathcal{S}$, $\forall t,q\in\widehat{\mathcal{T}}$. For handling this coupling, we define new optimization variable $\beta_{ s,t,p,q}=b_{s,t}b_{p,q}$. Since $b_{s,t}$ and $b_{p,q}$ are binary variables, we apply the big-M formulation to represent $\beta_{ s,t,p,q}=b_{s,t}b_{p,q}$ equivalently in terms of the following convex constraints
\begin{eqnarray}
&&\hspace*{-14mm}\mbox{C6a:~}0\leq \beta_{ s,t,p,q}\leq 1,\hspace*{5mm}
\mbox{C6b:~}\beta_{ s,t,p,q}\leq b_{s,t},\\
&&\hspace*{-14mm}\mbox{C6c:~}\beta_{ s,t,p,q}\leq b_{p,q},\hspace*{9mm}
\mbox{C6d:~}\beta_{ s,t,p,q}\geq b_{s,t}+b_{p,q}-1.
\end{eqnarray}
Then, we insert $\mathbf{h}_{\mathrm{I},k}$ and $\mathbf{h}_{\mathrm{E},j}$ defined in \eqref{channel1} back into \eqref{SINR} and \eqref{HarvestedPower} and rewrite the SINR of IR $k$ and the received RF power at ER $j$ as follows, respectively,
\begin{eqnarray}
\hspace*{-4mm}\Gamma_k&\hspace*{-1mm}=\hspace*{-1mm}&\frac{\underset{\substack{s\in\mathcal{S},t\in\widehat{\mathcal{T}}}}{\sum }\underset{\substack{p\in\mathcal{S},q\in\widehat{\mathcal{T}}}}{\sum }\beta_{ s,t,p,q}\mathbf{h}_{\mathrm{I},k,s,t}^H \mathbf{W}_k\mathbf{h}_{\mathrm{I},k,p,q}}{\underset{\substack{s\in\mathcal{S},t\in\widehat{\mathcal{T}}}}{\sum }\underset{\substack{p\in\mathcal{S},q\in\widehat{\mathcal{T}}}}{\sum }\beta_{ s,t,p,q}\mathbf{h}_{\mathrm{I},k,s,t}^H\left (  \underset{r\in\mathcal{K}\setminus \left \{ k \right \} }{\sum }\mathbf{W}_r+\mathbf{V}\right )\mathbf{h}_{\mathrm{I},k,p,q}+\sigma ^2_{\mathrm{I}_k}},\label{Gamma2}\\
\hspace*{-4mm}P_j^{\mathrm{ER}}&\hspace*{-1mm}=\hspace*{-1mm}&\underset{\substack{s\in\mathcal{S},t\in\widehat{\mathcal{T}}}}{\sum }\underset{\substack{p\in\mathcal{S},q\in\widehat{\mathcal{T}}}}{\sum }\beta_{ s,t,p,q}\mathbf{h}_{\mathrm{E},j,s,t}^H\left (  \underset{k\in\mathcal{K}}{\sum }\mathbf{W}_k+\mathbf{V}\right )\mathbf{h}_{\mathrm{E},j,p,q}.\label{Power2}
\end{eqnarray}
We note that there are still coupled optimization variables in \eqref{Gamma2} and \eqref{Power2}, i.e., $\beta_{ s,t,p,q}\mathbf{W}_k$ and $\beta_{s,t,p,q}\mathbf{V}$. To overcome this difficulty, we again apply the big-M formulation. In particular, we define new optimization variables $\widehat{\mathbf{W}}_{k,s,t,p,q}=\beta_{ s,t,p,q}\mathbf{W}_k$ and $\widehat{\mathbf{V}}_{s,t,p,q}=\beta_{s,t,p,q}\mathbf{V}$ to decompose the product terms by imposing the following additional convex constraints:
\begin{eqnarray}
\hspace*{-16mm}&&\mbox{C7a:~}\widehat{\mathbf{W}}_{k,s,t,p,q}\preceq\beta_{ s,t,p,q}P^{\mathrm{max}}\mathbf{I}_{N_{\mathrm{T}}},\\
\hspace*{-16mm}&&\mbox{C7b:~}\widehat{\mathbf{W}}_{k,s,t,p,q}\succeq \mathbf{W}_k-(1-\beta_{ s,t,p,q})P^{\mathrm{max}}\mathbf{I}_{N_{\mathrm{T}}},\\
\hspace*{-16mm}&&\mbox{C7c:~}\widehat{\mathbf{W}}_{k,s,t,p,q}\preceq \mathbf{W}_k,\hspace*{11mm}\mbox{C7d:~}\widehat{\mathbf{W}}_{k,s,t,p,q}\succeq \mathbf{0},\\
\hspace*{-16mm}&&\mbox{C8a:~}\widehat{\mathbf{V}}_{s,t,p,q}\preceq\beta_{ s,t,p,q}P^{\mathrm{max}}\mathbf{I}_{N_{\mathrm{T}}},\\
\hspace*{-16mm}&&\mbox{C8b:~}\widehat{\mathbf{V}}_{s,t,p,q}\succeq \mathbf{V}-(1-\beta_{ s,t,p,q})P^{\mathrm{max}}\mathbf{I}_{N_{\mathrm{T}}},\\
\hspace*{-16mm}&&\mbox{C8c:~}\widehat{\mathbf{V}}_{s,t,p,q}\preceq \mathbf{V},\hspace*{18mm}
\mbox{C8d:~}\widehat{\mathbf{V}}_{s,t,p,q}\succeq \mathbf{0},
\end{eqnarray}
where $P^{\mathrm{max}}$ denotes the maximum transmit power budget available at the BS. Then, constraints C1 and C2 can be respectively recast as follows
\begin{eqnarray}
\label{Chat1}
\hspace*{-12mm}&&\hspace*{-12mm}\widehat{\mbox{C1}}\mbox{:}\hspace*{1mm}\frac{1}{\Gamma_{\mathrm{req}_k}}\underset{\substack{s\in\mathcal{S},\\t\in\widehat{\mathcal{T}}}}{\sum }\underset{\substack{p\in\mathcal{S},\\q\in\widehat{\mathcal{T}}}}{\sum }\mathrm{Tr}\big(\mathbf{h}_{\mathrm{I},k,p,q}\mathbf{h}_{\mathrm{I},k,s,t}^H \widehat{\mathbf{W}}_{k,s,t,p,q}\big)\hspace*{-1mm}-\hspace*{-2mm}\underset{\substack{s\in\mathcal{S},\\t\in\widehat{\mathcal{T}}}}{\sum }\underset{\substack{p\in\mathcal{S},\\q\in\widehat{\mathcal{T}}}}{\sum }\mathrm{Tr}\Big (\mathbf{h}_{\mathrm{I},k,p,q}\mathbf{h}_{\mathrm{I},k,s,t}^H\big (\hspace*{-2mm}\underset{r\in\mathcal{K}\setminus \left \{ k \right \} }{\sum }\hspace*{-2mm}\widehat{\mathbf{W}}_{r,s,t,p,q}\hspace*{-1mm}+\hspace*{-1mm}\widehat{\mathbf{V}}_{s,t,p,q}\big )\Big )\hspace*{-1mm}\geq \hspace*{-1mm}\sigma ^2_{\mathrm{I}_k},\\[2mm]\notag\\
\hspace*{-12mm}&&\hspace*{-12mm}\widehat{\mbox{C2}}\mbox{:}\hspace*{1mm}C_{\mathrm{req}_j}\geq \mathrm{exp}\Big(-\varrho _j\underset{\substack{s\in\mathcal{S},\\t\in\widehat{\mathcal{T}}}}{\sum }\underset{\substack{p\in\mathcal{S},\\q\in\widehat{\mathcal{T}}}}{\sum }\mathrm{Tr}\big (\mathbf{h}_{\mathrm{E},j,p,q}\mathbf{h}_{\mathrm{E},j,s,t}^H(  \underset{k\in\mathcal{K}}{\sum }\widehat{\mathbf{W}}_{k,s,t,p,q}+\widehat{\mathbf{V}}_{s,t,p,q})\big ) \Big ),\forall j, \label{Chat2}
\end{eqnarray}
where constant $C_{\mathrm{req}_j}$ in $\widehat{\mbox{C2}}$ is defined as $C_{\mathrm{req}_j}=(\frac{a_j}{E_{\mathrm{req}_j}(1-\Xi_j)+a_j\Xi_j}-1)\mathrm{exp}(-\varrho _jc_j)$.

Then, we recast the optimization problem in \eqref{prob1} as follows
\begin{eqnarray}
\label{prob2}
&&\hspace*{-6mm}\underset{\substack{\mathbf{W}_k,\widehat{\mathbf{W}}_{k,s,t,p,q}\in\mathbb{H}^{N_{\mathrm{T}}},\\\mathbf{V},\widehat{\mathbf{V}}_{s,t,p,q}\in\mathbb{H}^{N_{\mathrm{T}}},\\b_{s,t},\beta_{ s,t,p,q}}}{\mino} \,\, \,\, \underset{\substack{s\in\mathcal{S},\\t\in\widehat{\mathcal{T}}}}{\sum }\underset{\substack{p\in\mathcal{S},\\q\in\widehat{\mathcal{T}}}}{\sum }\mathrm{Tr}\left (\underset{k\in\mathcal{K}}{\sum }\widehat{\mathbf{W}}_{k,s,t,p,q}+\widehat{\mathbf{V}}_{s,t,p,q}\right )\notag\\
&&\hspace*{-12mm}\mbox{s.t.~}\widehat{\mbox{C1}},\widehat{\mbox{C2}},\mbox{C3-C5},\mbox{C6a-C6d},\mbox{C7a-C7d},\mbox{C8a-C8d},\mbox{C9:~}\mathrm{Rank}(\mathbf{W}_k)\leq 1,~\forall k,
\end{eqnarray}
where $\mathbf{W}_k\in\mathbb{H}^{N_{\mathrm{T}}}$ and constraint C9 are imposed to guarantee that $\mathbf{W}_k=\mathbf{w}_k\mathbf{w}_k^H$ holds after optimization. We note that the binary constraint in C4 and rank-one constraint C9 are still obstacles to solving problem \eqref{prob2}. Nevertheless, in the next subsection, we develop a BnB-based algorithm to optimally solve \eqref{prob2}.

%%%%%%%%%%%%%%%%%%%%%%%%%%%%%%%%
\subsection{Optimal Resource Allocation Scheme}
The BnB approach is a promising systematic partial enumeration strategy to optimally solve discrete and combinatorial optimization problems. Given an optimization problem with a finite number of binary optimization variables, the BnB-based algorithm is guaranteed to find the globally optimal solution in a finite number of iterations \cite{horst2013global}. BnB algorithms have been widely adopted to optimally solve communication resource allocation problems involving binary variables such as optimal user scheduling \cite{7934461} and optimal subcarrier assignment \cite{5200968}. Thanks to the series of transformations applied in Section \ref{problem transformation}, the reformulated equivalent problem in \eqref{prob2} is in the canonical form that allows the application of the BnB concept to develop an optimal algorithm. The basic principle of BnB-based algorithms is to exploit a tree traversal where the feasible set of the main problem is mapped to the root. BnB-based algorithms explore all branches of the tree, where each node of the tree represents a subset of the solution set. For each node, a subproblem based on the corresponding subset is formulated and both an upper bound and a lower bound are constructed. These bounds are then utilized to check the optimality of a given subproblem. A node is discarded if it cannot produce a better solution than the current best solution found by the algorithm. Based on a pre-defined node selection strategy, the tree traversal proceeds by selecting and branching one node into two new nodes in each iteration of the BnB-based algorithm. As the tree structure continues to expand, the feasible set is progressively partitioned into smaller subsets and the current best solution is updated leading to improved objective values in the course of the iterations. Following the above procedure, the gap between the upper bound and the lower bound gradually vanishes in each iteration and the BnB-based algorithm converges to the globally optimal solution of the considered optimization problem. In this subsection, we present the construction of the bounds, the partition rule, and the branching strategy for the problem in \eqref{prob2}, and then develop the optimal resource allocation algorithm.
\subsubsection{Lower and Upper Bounds}
We denote the search space of the proposed BnB-based algorithm as $\mathcal{B}$, where $\mathcal{B}$ is the product of $ST$ binary sets. In particular, $\mathcal{B}$ is given by $\mathcal{B}=\underset{s\in\mathcal{S}}{\prod}\underset{t\in\mathcal{T}}{\prod}\mathcal{B}_{s,t}$, where $\mathcal{B}_{s,t}\overset{\Delta }{=}\left \{ 0,1 \right \}$, $\forall s$, $t$. Then, we define a continuous optimization variable $0\leq\widetilde{b}_{s,t}\leq1$ and rewrite optimization problem \eqref{prob2} as follows:
\begin{eqnarray}
\label{prob3}
&&\hspace*{-7mm}\underset{\substack{\mathbf{W}_k,\widetilde{\mathbf{W}}_{k,s,t,p,q}\in\mathbb{H}^{N_{\mathrm{T}}},\\\mathbf{V},\widetilde{\mathbf{V}}_{s,t,p,q}\in\mathbb{H}^{N_{\mathrm{T}}},\\\widetilde{b}_{s,t},~\widetilde{\beta}_{ s,t,p,q}}}{\mino} \,\, \,\, F_{\mathrm{L}}(\widetilde{b}_{s,t},\mathbf{W}_k,\mathbf{V})\overset{\Delta }{=}\underset{\substack{s\in\mathcal{S},\\t\in\widehat{\mathcal{T}}}}{\sum }\underset{\substack{p\in\mathcal{S},\\q\in\widehat{\mathcal{T}}}}{\sum }\mathrm{Tr}\left (\underset{k\in\mathcal{K}}{\sum }\widetilde{\mathbf{W}}_{k,s,t,p,q}+\widetilde{\mathbf{V}}_{s,t,p,q}\right )\notag\\
&&\hspace*{-12mm}\mbox{s.t.~}\mbox{C3},\mbox{C9},\notag\\
&&\hspace*{-5.5mm}\widetilde{\mbox{C1}}\mbox{:}\hspace*{2mm}\frac{1}{\Gamma_{\mathrm{req}_k}}\underset{\substack{s,p\in\mathcal{S},\\t,q\in\widetilde{\mathcal{T}}}}{\sum}\mathrm{Tr}\big(\mathbf{h}_{\mathrm{I},k,p,q}\mathbf{h}_{\mathrm{I},k,s,t}^H \widetilde{\mathbf{W}}_{k,s,t,p,q}\big)\notag\\
&&\hspace*{1.5mm}-\underset{\substack{s,p\in\mathcal{S},\\t,q\in\widetilde{\mathcal{T}}}}{\sum }\mathrm{Tr}\Big (\mathbf{h}_{\mathrm{I},k,p,q}\mathbf{h}_{\mathrm{I},k,s,t}^H \big (\underset{r\in\mathcal{K}\setminus \left \{ k \right \} }{\sum }\widetilde{\mathbf{W}}_{r,s,t,p,q}+\widetilde{\mathbf{V}}_{s,t,p,q}\big )\Big )\geq \sigma ^2_{\mathrm{I}_k},~\forall k,\notag\\
\hspace*{-5mm}&&\hspace*{-6mm}\widetilde{\mbox{C2}}\mbox{:}\hspace*{1mm}C_{\mathrm{req}_j}\geq \mathrm{exp}\Big(-\varrho _j\underset{\substack{s,p\in\mathcal{S},\\t,q\in\widetilde{\mathcal{T}}}}{\sum }\mathrm{Tr}\big ( \mathbf{h}_{\mathrm{E},j,p,q}\mathbf{h}_{\mathrm{E},j,s,t}^H(  \underset{k\in\mathcal{K}}{\sum }\widetilde{\mathbf{W}}_{k,s,t,p,q}+\widetilde{\mathbf{V}}_{s,t,p,q})\big ) \Big ),~\forall j,\notag\\
&&\hspace*{-5.5mm}\widetilde{\mbox{C4}}\mbox{:~}0\leq \widetilde{b}_{s,t}\leq 1,~\forall s,~t,\hspace*{8mm}\widetilde{\mbox{C5}}\mbox{:~}\underset{ s\in\mathcal{S}}{\sum}\widetilde{b}_{s,t}=1,~\forall t,\hspace*{8mm} \widetilde{\mbox{C6}}\mbox{a}\mbox{:~}0\leq \widetilde{\beta}_{ s,t,p,q}\leq 1,\notag\\
&&\hspace*{-5.5mm}\widetilde{\mbox{C6}}\mbox{b}\mbox{:~}\widetilde{\beta}_{ s,t,p,q}\leq \widetilde{b}_{s,t},
\hspace*{18mm}\widetilde{\mbox{C6}}\mbox{c}\mbox{:~}\widetilde{\beta}_{ s,t,p,q}\leq \widetilde{b}_{p,q},\hspace*{9mm}
\widetilde{\mbox{C6}}\mbox{d}\mbox{:~}\widetilde{\beta}_{ s,t,p,q}\geq \widetilde{b}_{s,t}+\widetilde{b}_{p,q}-1,\notag\\
&&\hspace*{-5.5mm}\widetilde{\mbox{C7}}\mbox{a}\mbox{:~}\widetilde{\mathbf{W}}_{k,s,t,p,q}\preceq\widetilde{\beta}_{ s,t,p,q}P^{\mathrm{max}}\mathbf{I}_{N_{\mathrm{T}}},\hspace*{10mm}\widetilde{\mbox{C7}}\mbox{b}\mbox{:~}\widetilde{\mathbf{W}}_{k,s,t,p,q}\succeq \mathbf{W}_k-(1-\widetilde{\beta}_{ s,t,p,q})P^{\mathrm{max}}\mathbf{I}_{N_{\mathrm{T}}},\notag\\
&&\hspace*{-5.5mm}\widetilde{\mbox{C7}}\mbox{c}\mbox{:~}\widetilde{\mathbf{W}}_{k,s,t,p,q}\preceq \mathbf{W}_k,\hspace*{6mm}\widetilde{\mbox{C7}}\mbox{d}\mbox{:~}\widetilde{\mathbf{W}}_{k,s,t,p,q}\succeq \mathbf{0},\hspace*{6mm}\widetilde{\mbox{C8}}\mbox{a}\mbox{:~}\widetilde{\mathbf{V}}_{s,t,p,q}\preceq\widetilde{\beta}_{ s,t,p,q}P^{\mathrm{max}}\mathbf{I}_{N_{\mathrm{T}}},\notag\\
&&\hspace*{-5.5mm}\widetilde{\mbox{C8}}\mbox{b}\mbox{:~}\widetilde{\mathbf{V}}_{s,t,p,q}\succeq \mathbf{V}-(1-\widetilde{\beta}_{ s,t,p,q})P^{\mathrm{max}}\mathbf{I}_{N_{\mathrm{T}}},\hspace*{6mm}\widetilde{\mbox{C8}}\mbox{c}\mbox{:~}\widetilde{\mathbf{V}}_{s,t,p,q}\preceq \mathbf{V},\hspace*{6mm}
\widetilde{\mbox{C8}}\mbox{d}\mbox{:~}\widetilde{\mathbf{V}}_{s,t,p,q}\succeq \mathbf{0},
\end{eqnarray}
where $\widetilde{\beta}_{ s,t,p,q}=\widetilde{b}_{s,t}\widetilde{b}_{p,q}$, $\widetilde{\mathbf{W}}_{k,s,t,p,q}=\widetilde{\beta}_{ s,t,p,q}\mathbf{W}_k$, and $\widetilde{\mathbf{V}}_{s,t,p,q}=\widetilde{\beta}_{ s,t,p,q}\mathbf{V}$. We note that constraint $\widetilde{\mbox{C4}}$ is a continuous relaxation of binary constraint C4. In general, solving the optimization problem in \eqref{prob3} may yield a non-binary solution. As a result, the optimal solution of the constraint-relaxed problem in \eqref{prob3}, i.e., $(\widetilde{b}_{s,t}^*,\mathbf{W}_k^*,\mathbf{V}^*)$, provides a lower bound for \eqref{prob2} which is denoted by $F_{\mathrm{L}}(\widetilde{b}_{s,t}^*,\mathbf{W}_k^*,\mathbf{V}^*)$. However, to optimally solve \eqref{prob3}, we still need to circumvent the non-convexity stemming from the unit-rank constraint C9. For handling this issue, we employ SDR and remove constraint C9. The rank-relaxed version of \eqref{prob2} is a convex optimization problem and can be efficiently solved by standard solvers such as CVX \cite{grant2008cvx}.
Next, we show the tightness of the relaxation by introducing the following theorem.
\par
\textit{Theorem 1:~}For given $\Gamma_{\mathrm{req}_k}>0$, the optimal solution $\mathbf{W}_k^*$ of the relaxed problem \eqref{prob3} always satisfies $\mathrm{Rank}(\mathbf{W}_k^*)=1$, $\forall k$.
\par
\textit{Proof:} Please refer to Appendix A. \qed
\par
On the other hand, we can also obtain an upper bound of \eqref{prob2} based on the solution produced by \eqref{prob3}. In particular, by relaxing the binary constraint and optimally solving the relaxed version of \eqref{prob3}, we obtain the optimal solution $\widetilde{b}_{s,t}^*$, where $0\leq\widetilde{b}_{s,t}^*\leq1$, $\forall s$, $t$. Then, based on $\widetilde{b}_{s,t}^*$, we construct a binary solution for the optimization problem in \eqref{prob2} by rounding each $\widetilde{b}_{s,t}^*$ to either $0$ or $1$. In particular, for $\forall t\in\mathcal{T}$, we round  the variable $\widetilde{b}_{s,t}^*$ with index $s^{\dagger}$ to $1$, where $s^{\dagger}$ is given by
%\begin{equation}
    %\overline{b}_{s,t}=\left \lfloor  \frac{\left \lceil 2\widetilde{b}_{s,t}^* \right \rceil}{2}\right \rfloor,~\forall s,t.\label{rounding}
%\end{equation}
\begin{equation}
\vspace*{-2mm}
    s^{\dagger}=\mathrm{arg}~\underset{s\in\mathcal{S}}{\mathrm{max~}}\widetilde{b}_{s,t}^*,~\forall t\in\mathcal{T}.\label{rounding}
\end{equation}
Then, we set all the other $\widetilde{b}_{s,t}^*$ to $0$, $\forall t\in\mathcal{T}$, and denote the rounded solution by $\overline{b}_{s,t}$. We note that given the optimal solution $\mathbf{W}_k^*$ and $\mathbf{V}^*$ of \eqref{prob3}, the rounded solution $\overline{b}_{s,t}$ may violate the constraints of \eqref{prob2} and cause infeasibility. Hence, we insert the rounded solution $\overline{b}_{s,t}$ back into \eqref{prob2} and solve the rank-relaxed version of \eqref{prob2} for the optimal solution $\mathbf{W}_k^{**}$ and $\mathbf{V}^{**}$.\footnote{Since the optimal solution of \eqref{prob3}, i.e., $\widetilde{b}_{s,t}^*$, and its rounded version $\overline{b}_{s,t}$ generally lead to different optimal beamforming policies, we denote the optimal beamforming policy associated with $\overline{b}_{s,t}$ by $(\mathbf{W}_k^{**},\mathbf{V}^{**})$ to avoid ambiguity.} Then, we can obtain the corresponding upper bound of the objective function value $F_{\mathrm{U}}(\overline{b}_{s,t},\mathbf{W}_k^{**},\mathbf{V}^{**})$. Now, we have acquired both a lower bound and an upper bound for the optimization problem in \eqref{prob2}.
%%%%%%%%%%%%%%%%%%%%%%%%%%%%%%%%%%%%%%%%
\subsubsection{Partitioning Rule and Branching Strategy}
%\footnote{To reduce the computational complexity, a node is removed from the search tree if its lower bound is larger than the upper bound of all the existing nodes.}
In each iteration of the BnB algorithm, we select a node in the search tree and branch the corresponding parent problem into two new subproblems, where we use superscript $(j)$ to denote the iteration index of the optimization variables. In particular, among all available nodes, we select the node associated with the smallest lower bound and partition its set according to the Euclidean distance between $\widetilde{b}_{s,t}$ and its rounded version $\overline{b}_{s,t}$. Specifically, in the $j$-th iteration, we branch the node with index $(s^*,t^*)$, where $(s^*,t^*)$ is given by
\begin{equation}
    (s^*,t^*)=\mathrm{arg}~\underset{s,t}{\mathrm{max}}\left | \widetilde{b}_{s,t}^{(j)}-\overline{b}_{s,t}^{(j)} \right |. \label{partition}
\end{equation}
Accordingly, the feasible set of $b_{s^*,t^*}$, i.e., $\mathcal{B}_{s^*,t^*}^{(j)}$, is further divided into two new subsets $(\mathcal{B}^{(j)}_{s^*,t^*})_l$ and $(\mathcal{B}^{(j)}_{s^*,t^*})_r$, which are associated with $b_{s^*,t^*}^{(j)}=0$ and $b_{s^*,t^*}^{(j)}=1$, respectively.
Then, in the $j$-th iteration, we focus on the following two subproblems $\mathcal{P}_i$
\begin{eqnarray}
\label{prob4}
\mathcal{P}_i:\hspace*{12mm}&&\hspace*{-12mm}\underset{\substack{\mathbf{W}_k,\widehat{\mathbf{W}}_{k,s,t,p,q}\in\mathbb{H}^{N_{\mathrm{T}}},\\\mathbf{V},\widehat{\mathbf{V}}_{s,t,p,q}\in\mathbb{H}^{N_{\mathrm{T}}},\\b_{s,t},\beta_{ s,t,p,q}}}{\mino} \,\, \,\, \underset{\substack{s\in\mathcal{S},\\t\in\widehat{\mathcal{T}}}}{\sum }\underset{\substack{p\in\mathcal{S},\\q\in\widehat{\mathcal{T}}}}{\sum }\mathrm{Tr}\left (\underset{k\in\mathcal{K}}{\sum }\widehat{\mathbf{W}}_{k,s,t,p,q}+\widehat{\mathbf{V}}_{s,t,p,q}\right )\notag\\
&&\hspace*{-12mm}\mbox{s.t.~}\mbox{C1},\mbox{C2},\mbox{C3},\mbox{C5},\mbox{C6}\mbox{a}\mbox{-}\mbox{C6}\mbox{d},\mbox{C7}\mbox{a}\mbox{-}\mbox{C7}\mbox{d},\mbox{C8}\mbox{a}\mbox{-}\mbox{C8}\mbox{d},\notag\\
&&\hspace*{-5.5mm}\overline{\mbox{C4}}\mbox{:~} b_{s,t}^{(j)}\in\mathcal{B}^{(j)}_{s,t},~\forall s\in\mathcal{S}\setminus \left \{ s^* \right \},~t\in\mathcal{T}\setminus \left \{ t^* \right \},\notag\\
&&\hspace*{-5.5mm}\mbox{C10}\mbox{:~}b_{s^*,t^*}^{(j)}=i,
\end{eqnarray}
where $i\in\left \{0,1\right \}$. We note that constraint $\overline{\mbox{C4}}$ contains both the transmission mode selection variables determined in the previous iterations and the undetermined binary optimization variables to be optimized in the future iterations, which makes \eqref{prob4} a non-convex optimization problem. By relaxing the undetermined binary $b_{s,t}$ to continuous optimization variables in $[0,1]$, we solve a relaxed version of \eqref{prob4} to obtain the optimal solution and the corresponding objective function value. Then, based on the optimal solution, we determine the rounded solution according to \eqref{rounding}. Subsequently, we solve \eqref{prob2} by inserting the rounded solution and compute the corresponding objective function value. Based on these objective function values, we can respectively update the upper bound and lower bound in each iteration. In Figure \ref{BnBfigure}, we provide an example for the BnB search tree for $T=2$ and $S=2$. Firstly, the root node branches into two new nodes associated with $b_{s,t}=0$ and $b_{s,t}=1$. In each iteration, the search tree is expanded by adding two new nodes while it is pruned by discarding those nodes (grey dots) that are worse than the current upper bound. We note that the branching procedure is exhaustive due to the limited depth of the search tree and the finite number of nodes at each depth. As a result, the BnB algorithm always terminates within a finite number of iterations.
%%%%%%%%%%%%%%%%%%%%%%%%%%%%%%%%%%%%%%%%
%\begin{figure}[t]\vspace*{0mm}
 %\centering
%\includegraphics[width=2.4in]{BnB_simple.eps}
%\vspace*{-2mm}
%\caption{An illustration of the BnB search tree for $S=2$ and $T=2$. The green arrows and dots correspond to the link to the optimal node (yellow star). The red and grey dots correspond to non-optimal feasible nodes and discarded nodes, respectively.}\vspace*{-2mm}\label{BnBfigure}
%\caption{An illustration of a BnB search tree with $S=2$ and $T=2$. The green arrows and dots comprise the search path to the optimal node (yellow star). The red, grey, and orange dots denote the non-optimal feasible nodes and the nodes discarded due to infeasibility and bound (whose lower bound is larger than the upper bound of all existing nodes), respectively.}
%\end{figure}
%%%%%%%%%%%%%%%%%%%%%%%%%%%%%%%%%%%%%%%%
%The searching space $\mathcal{B}$ of the BnB-based algorithm is the product of $S(T+1)$ binary sets, i.e., $\underset{\substack{s\in\mathcal{S},\\t\in\widehat{\mathcal{T}}}}{\prod}\mathcal{B}_{s,t}$, where $\mathcal{B}_{s,t}\overset{\Delta }{=}\left \{ 0,1 \right \}$, $\forall s$, $t$. 
\begin{algorithm}[t]
\caption{BnB-based Algorithm}
\begin{algorithmic}[1]
\small
\STATE  Solve \eqref{prob3} to obtain optimal solution $( \widetilde{b}_{s,t}^*)^{(1)}$ and compute lower bound $L^{(1)}=F_{\mathrm{L}}\big((\widetilde{b}_{s,t}^*)^{(1)},(\mathbf{W}_k^*)^{(1)},(\mathbf{V}^*)^{(1)}\big)$. Compute the rounded binary solution $(\overline{b}_{s,t})^{(1)}$ according to \eqref{rounding} and obtain the corresponding upper bound $U^{(1)}=F_{\mathrm{U}}\big((\overline{b}_{s,t})^{(1)},(\mathbf{W}_k^{**})^{(1)},(\mathbf{V}^{**})^{(1)}\big)$. Initialize the search tree $\mathcal{T}_{\mathrm{BnB}}$ by adding the root node
associated with $\mathcal{B}^{(1)}$ and $( \widetilde{b}_{s,t}^* )^{(1)}$. Set convergence tolerance $0<\varepsilon_{\mathrm{BnB}}\ll1$ and iteration index $j=1$.
\REPEAT
\STATE Select the node corresponding to the smallest lower bound $F_{\mathrm{L}}\big((\widetilde{b}_{s,t}^*)^{(j)},(\mathbf{W}_k^*)^{(j)},(\mathbf{V}^*)^{(j)}\big)$
\STATE Partition the feasible set associated with the selected
node into two subsets $(\mathcal{B}_{s^*,t^*})^{(j)}_l$ and $(\mathcal{B}_{s^*,t^*})^{(j)}_r$ according to \eqref{partition}
\STATE Solve the relaxed version of the two subproblems $\mathcal{P}_0$ and $\mathcal{P}_1$ in \eqref{prob4} to obtain optimal solutions $(\widetilde{b}_{s^*,t^*}=0,  \widetilde{b}_{s,t}^*)^{(j)}_l$ and $( \widetilde{b}_{s^*,t^*}=1,  \widetilde{b}_{s,t}^*)^{(j)}_r$, $\forall s\in\mathcal{S}\setminus \left \{ s^* \right \},~t\in\mathcal{T}\setminus \left \{ t^* \right \}$, and store the corresponding objective function values
\STATE Compute the rounded solutions based on $\left\{(\widetilde{b}_{s^*,t^*}=0,  \widetilde{b}_{s,t}^*)^{(j)}_l, ( \widetilde{b}_{s^*,t^*}=1,  \widetilde{b}_{s,t}^*)^{(j)}_r\right\}$ and obtain $\left\{(\overline{b}_{s,t})^{(j)}_l, (\overline{b}_{s,t})^{(j)}_r\right\}$
\STATE Solve the problem in \eqref{prob2} based on $(\overline{b}_{s,t})^{(j)}_l$ and $(\overline{b}_{s,t})^{(j)}_r$ and store the corresponding objective function values
\STATE Expand the tree $\mathcal{T}_{\mathrm{BnB}}$ by adding the two new nodes associated with $(\mathcal{B}_{s^*,t^*})^{(j)}_l$ and $(\widetilde{b}_{s^*,t^*}=0,  \widetilde{b}_{s,t}^*)^{(j)}_l$ and $(\mathcal{B}_{s^*,t^*})^{(j)}_r$ and $( \widetilde{b}_{s^*,t^*}=1,  \widetilde{b}_{s,t}^*)^{(j)}_r$, respectively
\STATE Among all existing nodes in $\mathcal{T}_{\mathrm{BnB}}$, update $L^{(j)}$ and $U^{(j)}$ as the smallest upper bound $F_{\mathrm{U}}\big((\overline{b}_{s,t})^{(j)},(\mathbf{W}_k^{**})^{(j)},(\mathbf{V}^{**})^{(j)}\big)$ and lower bound $F_{\mathrm{L}}\big((\widetilde{b}_{s,t}^*)^{(j)},(\mathbf{W}_k^*)^{(j)},(\mathbf{V}^*)^{(j)}\big)$, respectively
\STATE Set $j=j+1$
\UNTIL $\frac{U^{(j-1)}-L^{(j-1)}}{L^{(j-1)}}\leq \varepsilon_{\mathrm{BnB}}$
\STATE Output the optimal solution $(\overline{b}_{s,t})^{(j-1)}$ and the corresponding beamforming policy $\big((\mathbf{W}_k^{**})^{(j-1)},(\mathbf{V}^{**})^{(j-1)}\big)$
\end{algorithmic}
\end{algorithm}
\par
The BnB-based algorithm for optimally solving optimization problem in \eqref{prob2} is summarized in \textbf{Algorithm 1}. In each iteration of the BnB-based algorithm, we denote the sets and the solutions associated with the two partitioned child nodes by subscripts $l$ and $r$, respectively, to distinguish them from those associated with the parent node. The aforementioned set partitioning, node branching, and bound updating steps are repeatedly performed until the difference between the lower bound and the upper bound of \eqref{prob2} is less than a pre-defined convergence tolerance factor. It is known that BnB-based algorithms are guaranteed to converge to an $\varepsilon_{\mathrm{BnB}}$-optimal solution within a finite number of iterations \cite{horst2013global}, where $\varepsilon_{\mathrm{BnB}}$ is the maximum error tolerance. The proof of convergence for the adopted BnB algorithm follows directly from \cite{maranas1997global}. The developed BnB-based algorithm can serve as a performance benchmark for any suboptimal algorithm. However, the computational complexity of the BnB-based algorithm scales exponentially with the number of transmission modes $S$ and the number of tiles $T$. To strike a balance between optimality and computational complexity, in the next subsection, we develop an SCA-based algorithm which determines a suboptimal solution of the considered optimization problem in polynomial time.
\par

%\footnote{We note that if we set the number of tiles as the number of IRS elements and map the discrete phase shift set to the transmission mode set, the proposed BnB-based algorithm can still serve as a benchmark for any suboptimal scheme developed under element-wise optimization framework.}

\subsection{Suboptimal Resource Allocation Scheme}
We start with the non-convex optimization problem in \eqref{prob2}. To facilitate efficient resource allocation algorithm design, we first rewrite constraint C4 equivalently as follows:
\begin{eqnarray}
\mbox{C4a:~}\underset{\substack{s\in\mathcal{S},t\in\widehat{\mathcal{T}}} }{\sum }b_{s,t}-b_{s,t}^2\leq 0~\mbox{and}~
\mbox{C4b:~}0\leq b_{s,t}\leq 1,~\forall s,t.
\end{eqnarray}
We note that constraint $\mbox{C4a}$ involves a difference of convex (d.c.) functions and hence is still non-convex with respect to $b_{s,t}$. To circumvent this obstacle, we employ the penalty method \cite{ben1997penalty} and recast \eqref{prob2} as follows:
\begin{eqnarray}
\label{prob5}
&&\hspace*{-7mm}\underset{\substack{\mathbf{W}_k,\widehat{\mathbf{W}}_{k,s,t,p,q}\in\mathbb{H}^{N_{\mathrm{T}}},\\\mathbf{V},\widehat{\mathbf{V}}_{s,t,p,q}\in\mathbb{H}^{N_{\mathrm{T}}},\\b_{s,t},\beta_{ s,t,p,q}}}{\mino} \,\, \,\, \underset{\substack{s\in\mathcal{S},\\t\in\widehat{\mathcal{T}}}}{\sum }\underset{\substack{p\in\mathcal{S},\\q\in\widehat{\mathcal{T}}}}{\sum }\mathrm{Tr}\left (\underset{k\in\mathcal{K}}{\sum}\widehat{\mathbf{W}}_{k,s,t,p,q}+\widehat{\mathbf{V}}_{s,t,p,q}\right )+\chi\underset{\substack{s\in\mathcal{S},\\t\in\widehat{\mathcal{T}}} }{\sum }(b_{s,t}-b_{s,t}^2)\notag\\
&&\hspace*{4mm}\mbox{s.t.}\hspace*{2mm}\widehat{\mbox{C1}},\widehat{\mbox{C2}},\mbox{C3},\mbox{C4b},\mbox{C5},\mbox{C6a-C6d},\mbox{C7a-C7d},\mbox{C8a-C8d},\mbox{C9},
\end{eqnarray}
where $\chi\gg0$ is a constant penalty factor which ensures that $b_{s,t}$ is binary. Next, we reveal the equivalence between problem \eqref{prob5} and problem \eqref{prob2} in the following theorem \cite{ben1997penalty}.
\par
\textit{Theorem 2:} Denote the optimal solution of problem \eqref{prob5} as $(b_{s,t})_i$ with penalty factor $\chi=\chi_i$. When $\chi_i$ is sufficiently large, i.e., $\chi=\chi_i\rightarrow \infty$, every limit point $(b_{s,t})$ of the sequence $\left \{ (b_{s,t})_i \right \}$ is an optimal solution of problem \eqref{prob2}.
\par
\textit{Proof:} The optimization problem in \eqref{prob5} has a similar structure as \cite[Problem (27)]{yu2019robust} and Theorem 2 can be proved following the same steps
as in \cite[Appendix C]{yu2019robust}. Due to the limited space, we omit the detailed proof of Theorem 2 for brevity.  \qed
\par 
We note that the objective function of \eqref{prob5} is in the canonical form of a difference of convex programming problem, which facilitates the application of SCA. In particular, for a given feasible point $b_{s,t}^{(m)}$ found in the $m$-th iteration of the SCA procedure, we construct a global underestimator of $b_{s,t}^2$ as follows
\begin{equation}
    b_{s,t}^2 \geq 2 b_{s,t}b_{s,t}^{(m)}-(b_{s,t}^{(m)})^2,~\forall s,t.
\end{equation}
\par
The optimization problem solved in the $(m+1)$-th iteration of the proposed algorithm is given by
\begin{eqnarray}
\label{prob6}
&&\hspace*{-7mm}\underset{\substack{\mathbf{W}_k,\widehat{\mathbf{W}}_{k,s,t,p,q}\in\mathbb{H}^{N_{\mathrm{T}}},\\\mathbf{V},\widehat{\mathbf{V}}_{s,t,p,q}\in\mathbb{H}^{N_{\mathrm{T}}},\\b_{s,t},\beta_{s,t,p,q}}}{\mino} \,\, \,\, f(\mathbf{W}_k,\mathbf{V},b_{s,t})\notag\\
&&\hspace*{2mm}\mbox{s.t.}\hspace*{2mm}\widehat{\mbox{C1}},\widehat{\mbox{C2}},\mbox{C3},\mbox{C4b},\mbox{C5-C9},
\end{eqnarray}
where $f(\mathbf{W}_k,\mathbf{V},b_{s,t})$ is defined as
\begin{equation}
    f(\mathbf{W}_k,\mathbf{V},b_{s,t})=\underset{\substack{s\in\mathcal{S},t\in\widehat{\mathcal{T}}}}{\sum }\underset{\substack{p\in\mathcal{S},q\in\widehat{\mathcal{T}}}}{\sum }\mathrm{Tr}\left (\underset{k\in\mathcal{K}}{\sum}\widehat{\mathbf{W}}_{k,s,t,p,q}+\widehat{\mathbf{V}}_{s,t,p,q}\right)+\chi\underset{\substack{s\in\mathcal{S},t\in\widehat{\mathcal{T}}}}{\sum }\Big(b_{s,t}-2 b_{s,t}b_{s,t}^{(m)}+(b_{s,t}^{(m)})^2\Big).
\end{equation}
We note that the only non-convex constraint in \eqref{prob6} is the unit-rank constraint C9. To overcome this, we omit constraint C9 by applying SDR. Following similar steps as in the Appendix, we can prove that the solution of the relaxed problem yields a rank one beamforming matrix. As a result, the rank-relaxed version of \eqref{prob6} becomes a standard convex optimization problem which can be solved by convex program solvers such as CVX \cite{grant2008cvx}. The overall algorithm is summarized in \textbf{Algorithm 2}. In each iteration of \textbf{Algorithm 2}, the objective function in \eqref{prob6} is monotonically decreasing. Moreover, as $\chi\rightarrow\infty $, the proposed algorithm asymptotically converges to a locally optimal solution of \eqref{prob2} in polynomial time. 
\par
\begin{algorithm}[t]
\caption{SCA-based Algorithm}
\begin{algorithmic}[1]
\small
\STATE Set initial point $\mathbf{W}_k^{(1)}$, $\mathbf{V}^{(1)}$, $\widehat{\mathbf{W}}_{k,s,t,p,q}^{(1)}$, $\widehat{\mathbf{V}}_{s,t,p,q}^{(1)}$, $b_{s,t}^{(1)}$, $\beta_{ s,t,p,q}^{(1)}$, iteration index $m=1$, and convergence tolerance $0<\varepsilon_{\mathrm{SCA}}\ll1$.
\REPEAT
\STATE For given $\mathbf{W}_k^{(m)}$, $\mathbf{V}^{(m)}$, $\widehat{\mathbf{W}}_{k,s,t,p,q}^{(m)}$, $\widehat{\mathbf{V}}_{s,t,p,q}^{(m)}$, $b_{s,t}^{(m)}$, $\beta_{ s,t,p,q}^{(m)}$ obtain an intermediate solution $\mathbf{W}_k^{(m+1)}$, $\mathbf{V}^{(m+1)}$, $\widehat{\mathbf{W}}_{k,s,t,p,q}^{(m+1)}$, $\widehat{\mathbf{V}}_{s,t,p,q}^{(m+1)}$, $b_{s,t}^{(m+1)}$, $\beta_{ s,t,p,q}^{(m+1)}$ by solving the relaxed version of problem \eqref{prob6}
\STATE Set $m=m+1$
\UNTIL $\frac{f(\mathbf{W}_k^{(m-1)},\mathbf{V}^{(m-1)},b_{s,t}^{(m-1)})-f(\mathbf{W}_k^{(m)},\mathbf{V}^{(m)},b_{s,t}^{(m)})}{f(\mathbf{W}_k^{(m)},\mathbf{V}^{(m)},b_{s,t}^{(m)})}\leq \varepsilon_{\mathrm{SCA}}$
\end{algorithmic}
\end{algorithm}
\begin{Remark}
In the literature, the commonly adopted optimization framework for IRSs aims at jointly optimizing the IRS elements over the entire phase shift domain \cite{8811733} or a given discrete phase shift set \cite{8930608}. Some advanced algorithms based on AO \cite{pan2020intelligent}, inner approximation (IA) \cite{yu2020irs}, and SCA \cite{wu2021intelligent} have been developed to tackle IRS optimization problems. Nevertheless, the computational complexity of these algorithms is typically proportional to at least the cubic power of the number of IRS elements. For a practically large IRS which usually comprises $M\geq500$ phase shift elements, the online element-wise IRS optimization design becomes prohibitive. In contrast, by adopting the proposed TT-based optimization framework and employing \textbf{Algorithm 2}, the computational complexity of IRS optimization scales with the number of tiles and the size of the refined transmission mode set specified in Section \ref{Pre-selection}, i.e., $T$ and $S$, instead of the number of IRS elements $M$. In particular, the (worst case) per iteration computational complexity of \textbf{Algorithm 2} is given by $\mathcal{O}\Big((K+1)S^2T^2N_{\mathrm{T}}^{3}+\big((K+1)S^2T^2\big)^2N_{\mathrm{T}}^{2}+\big((K+1)S^2T^2\big)^3\Big)$, where $\mathcal{O}\left ( \cdot  \right )$ is the big-O notation \cite[Theorem 3.12]{polik2010interior}. Hence, by properly adjusting the number of tiles and the number of transmission modes, the computational complexity of the developed algorithm becomes affordable for online optimization of large IRSs. Furthermore, as will be shown in the next section, the proposed scalable optimization framework enables us to flexibly adjust the trade-off between the computational complexity of the algorithm and the performance of IRS-assisted communication systems.
\end{Remark}

\begin{table}[t]
	\centering
	\begin{minipage}[t]{0.47\linewidth}\vspace*{10mm}  
		\centering
		\includegraphics[width=2.4in]{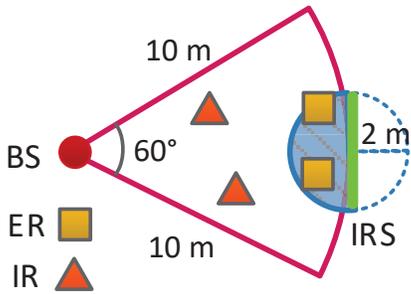}
        \captionof{figure}{\vspace*{-2mm}Simulation setup for an IRS-assisted SWIPT system, which consists of $K=2$ IRs and $J=2$ ERs.}
\label{setup}
	\end{minipage}\quad
	\begin{minipage}[t]{0.47\linewidth}%\vspace*{2mm} 
		\caption{\vspace*{-2mm}Simulation parameters\vspace*{-5mm}}
		\centering
\begin{tabular}{|l|l|}
\hline
    \hspace*{-1mm} Carrier center frequency, $f_c$ & $2.4$ GHz \\
\hline
    \hspace*{-1mm} Bandwidth, $B$ & $200$ kHz \\
\hline
    %\hspace*{-1mm} Noise power density, %$N_0$ & $-154$ dBm/Hz \\
%\hline
    %\hspace*{-1mm}$\alpha$ & Path loss exponent & $2$ \\
%\hline
    %\hspace*{-1mm}$N_{\mathrm{T}}$ & Number of antennas at the BS & $8$ \\
%\hline
    \hspace*{-1mm} Min. required SINR at each IR, $\Gamma_{\mathrm{req}}$ & $10$ dB \\ 
\hline
    \hspace*{-1mm} Min. required energy at each ER, $E_{\mathrm{req}}$ & $10$ $\mu$W \\ 
\hline
    \hspace*{-1mm} BS antenna gain, $G_i$ & $10$ dBi \\
%\hline
    %\hspace*{-1mm} EH parameter, $a_j$ & $20$ mW \cite{boshkovska2015practical}\\
\hline
    \hspace*{-1mm} EH parameters, $a_j$, $c_j$, $\varrho_j$ & $20$, $6400$, $0.003$\\
\hline
    \hspace*{-1mm} Numbers of scatterers in each link & $6$ \\
\hline
    \hspace*{-1mm} Noise power at receivers, $\sigma^2_{\mathrm{I}_k}$, $\sigma^2_{\mathrm{E}_j}$ & $-100$ dBm \\
\hline
    \hspace*{-1mm} Convergence tolerance, $\varepsilon_{\mathrm{BnB}}$, $\varepsilon_{\mathrm{SCA}}$ & $10^{-2}$ \\
\hline
    \hspace*{-1mm} Penalty factor, $\chi$ & $10^{3}$\\
\hline
\end{tabular}
\label{tab:parameters}
	\end{minipage}\vspace*{-8mm}
\end{table}
%%%%%%%%%%%%%%%%%%%%%%%%%%%%%%%%%%%%%%%
%\footnote{In this paper, we employ the readily available offline codebook design from \cite{najafi2020physics}. However, we note that one may develop an offline codebook that is specifically designed for the setup considered here (e.g., accounting for users in both near-field and far-field of the IRS) and hence achieve a higher performance. The design of such offline codebooks is beyond the scope of this paper and constitutes an interesting problem for future work. Nevertheless, we emphasize that the proposed online optimization algorithm is applicable to any adopted offline codebook.}
\section{Simulation Results}
\label{Large_IRS_simulation}
In this section, we evaluate the performance of the
proposed resource allocation schemes via simulations.
%%%%%%%%%%%%%%%%%%%%%%%%%%%%%%%%%%%%%%%%%%%%%%%%%
\subsection{Simulation Setup}
Figure \ref{setup} illustrates the schematic of the simulated multiuser MISO SWIPT system. We focus on the resource allocation algorithm design for one sector of a cell with a radius of $10$ m. The BS is equipped with $N_{\mathrm{T}}=8$ antennas, unless otherwise specified. We assume that there are $K=2$ IRs randomly and uniformly distributed in the considered SWIPT system. To enhance the system performance, we consider a rectangular IRS comprising $M=600$ elements. The IRS is located at the edge of the sector and is $10$ m away from the BS. There are $J=2$ ERs randomly and uniformly distributed within the charging zone between the BS and the IRS, which is a semicircular area (blue area) with the IRS at its center and a radius of $2$ m, cf. Figure \ref{setup}. To facilitate computationally efficient resource allocation algorithm design, we partition the $600$ IRS elements into $T$ tiles of equal size, where each tile comprises $600/T$ phase shift elements. We jointly design the elements of each tile offline to generate a set of transmission modes. Following a similar approach as in \cite[Section III-A]{najafi2020physics}, for all tiles, we generate a transmission mode set offline which is the product of a reflection codebook with $121$ elements and a wavefront phase codebook with $2$ elements\footnote{The reflection codebook enables the tile to reflect an incident signal with the desired phase shift, while the wavefront phase codebook facilitates the combination of the signals that arrive from different tiles at the receivers in a constructive or destructive manner.}. Then, we employ the three mode pre-selection criteria proposed in Section \ref{Pre-selection} and obtain the corresponding refined transmission mode sets. For a fair comparison, we adjust parameters $\delta_1$, $\delta_{2,r,i}$, and $\omega$ such that the refined transmission mode sets for all criteria have the same size $S$. In the following, unless otherwise specified, we pre-select the transmission modes based on Criterion 1. Moreover, we assume that the channel coefficients contained in $\mathbf{C}_{\mathrm{T}}$, $\mathbf{C}_{\mathrm{R}_k}$, and $\mathbf{C}_{\mathrm{D}_k}$ are impaired by free space path loss, shadowing, and Rayleigh fading. The path loss exponent is assumed to be $2$ for all channels while the path loss at a reference distance of $1$ m is set as $(\frac{c}{4\pi f_c})^2=40$ dB \cite{yu2019robust}. Assuming the direct links are severely shadowed, the shadowing attenuations are $-30$ dB and $0$ dB for the direct links and the reflected links, respectively. The AoAs and AoDs at the BS and the IRS are uniformly distributed random variables and are generated as follows: the azimuth angles and polarizations of the incident signal are uniformly distributed in the interval $[0,2\pi]$. The elevation angles of the IRS and the BS are uniformly distributed in the range of $[0,\pi/4]$ while the elevation angles of all users are uniformly distributed in the interval $[0,\pi]$. The simulation results shown in this section have been averaged over different channel realizations and the adopted parameter values are listed in Table \ref{tab:parameters}.
\par
\subsection{Baseline Schemes}
To investigate the effectiveness of the algorithms developed in this paper, we consider three baseline schemes. For baseline scheme 1, a transmission mode is randomly chosen from the refined transmission mode set and is assigned to each tile while the BS adopts an isotropic radiation pattern for $\mathbf{V}$. Then, we optimize the beamforming vector $\mathbf{w}_k$ and the power allocated to the covariance matrix of the energy signal $\mathbf{V}$ for minimization of the total transmit power. For baseline scheme 2, the IRS employs random phase shifts. Then, we jointly optimize the beamforming vector $\mathbf{w}_k$ and the covariance matrix of the energy signal $\mathbf{V}$ for minimization of the BS total transmit power. For baseline scheme 3, for each tile, we select the transmission mode corresponding to the channel vector with the largest Euclidean norm directly from the offline transmission mode set, and the BS employs maximum ratio transmission with respect to the corresponding channel vector, i.e., $\mathbf{w_\mathit{k}}=\dfrac{\sqrt{p_k}\mathbf{h}_{\mathrm{I},k}} {\left \| \mathbf{h}_{\mathrm{I},k}\right \|_2}$, where $p_k$ is the power allocated to IR $k$. Then, the transmit power at the BS is minimized by optimizing the power allocated to each IR and the covariance matrix of the energy signal.
%We consider two baseline schemes for comparison. For baseline scheme 1, we adopt zero-forcing (ZF) beamforming at the secondary BS and generate the phases of the IRS in a random manner. In particular, we fix the direction of beamformer $\mathbf{w}_k$ for desired user $k$ such that it lies in the null spaces of all the other users’ channels. Then, we solve a problem similar to problem \eqref{prob1} where we also optimize the power allocated to SU $k$, i.e., $p_k\in\mathbb{R}$. For baseline scheme 2, we assume that an IRS is not deployed. Then, by applying SCA, we optimize the beamforming vector $\mathbf{w}_k$ for maximization of the system sum rate based on problem \eqref{prob2}.
%%%%%%%%%%%%%%%%%%%%%%%%%%%%%%%%%%%%%%%%%%%%%%%%%%
\subsection{Convergence of the Proposed Algorithms}
%%%%%%%%%%%%%%%%%%%%%%%%%%%%%%%%%%%%%%%%%%%%%%%%%%%%%%%%
\begin{figure}[t]
\centering
%\vspace*{-2mm}
\hspace*{-4mm}\begin{minipage}[b]{0.47\linewidth}
    \centering
\hspace*{-2mm}\includegraphics[width=3.3in]{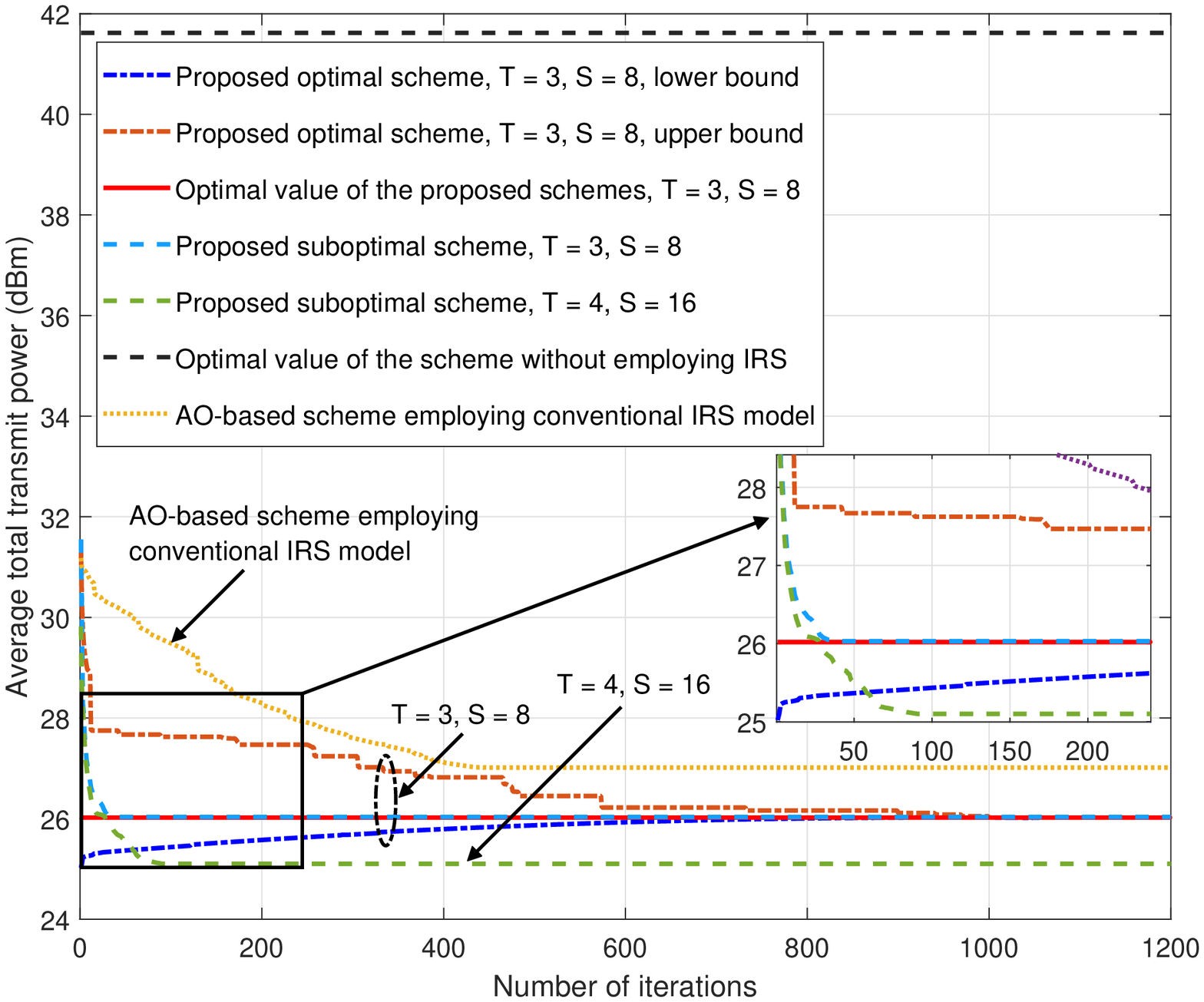}
\vspace*{-6mm}
\caption{Convergence behavior of different algorithms for $K=J=2$, $N_{\mathrm{T}}=10$, $\varepsilon_{\mathrm{BnB}}=10^{-4}$, $\varepsilon_{\mathrm{SCA}}=10^{-4}$, $\Gamma_{\mathrm{req}}=10$ dB, and $E_{\mathrm{req}}=5$ $\mu$W.}\label{poweriteraation_SWIPT}
\end{minipage}
\hspace*{4mm}\begin{minipage}[b]{0.47\linewidth}
    \centering 
\hspace*{-2mm}\includegraphics[width=3.3in]{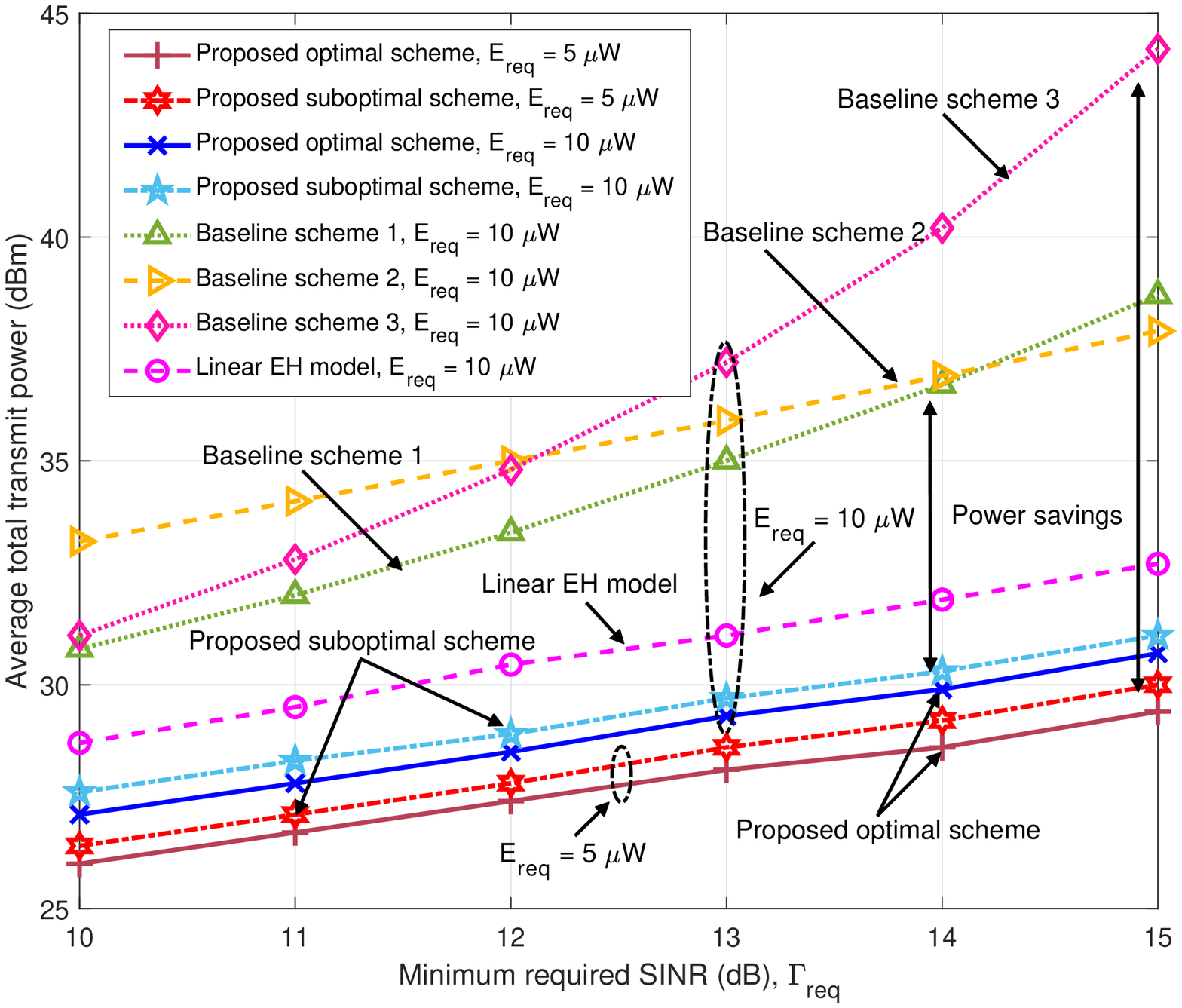}
\vspace*{-6mm}
\caption{Average total transmit power (dBm) versus minimum required SINR at IRs for different schemes with $K=J=2$, $N_{\mathrm{T}}=8$, $T=3$, $S=8$, $\Gamma_{\mathrm{req}}=10$ dB, and $E_{\mathrm{req}}=10$ $\mu$W.}\label{powerSINR_SWIPT}
\end{minipage}\vspace*{-8mm}
\end{figure}
%%%%%%%%%%%%%%%%%%%%%%%%%%%%%%%%%%%%%%%%%%%%%%%%%%
%\begin{figure}[t]\vspace*{0mm}
 %\centering
%\includegraphics[width=3.8in]{Large_cong_journal.eps}
%\vspace*{-4mm}
%\caption{Convergence for the proposed algorithm of different IRS models for $K=2$, $J=2$, $N_{\mathrm{T}}=8$, $\varepsilon_{\mathrm{SCA}}=10^{-4}$, $\Gamma_{\mathrm{req}_k}=10$ dB, and $E_{\mathrm{req}_j}=10$ $\mu$W.}\vspace*{-2mm}\label{poweriteraation_SWIPT}
%\end{figure}
In Figure \ref{poweriteraation_SWIPT}, we investigate the convergence behavior of the proposed algorithms for different IRS models. For $T=3$ and $S=8$, we observe that the upper bound and lower bound of the proposed optimal scheme monotonically converge to the same objective function value confirming the optimality of the proposed optimal scheme. Yet, the average number of iterations needed for achieving convergence is around $1$,$000$. This is due to the fact that the computational complexity of the proposed optimal algorithm increases exponentially with $T$ and $S$. Moreover, Figure \ref{poweriteraation_SWIPT} also confirms that the proposed suboptimal scheme achieves a close-to-optimal performance while enjoying a polynomial time computational complexity. For the suboptimal scheme, we also consider an IRS with $T=4$ and $S=16$. As we increase the number of tiles and enlarge the size of the refined transmission mode set (from $T=3$ and $S=8$ to $T=4$ and $S=16$), the total transmit power required for the proposed suboptimal scheme reduces by roughly $1$ dB. Correspondingly, the suboptimal scheme requires approximately $40$ additional iterations to converge for $T=4$ and $S=16$ than for $T=3$ and $S=8$. This confirms that by reconfiguring the tiles and resizing the refined transmission mode set, we can flexibly manipulate the trade-off between the system performance and the number of iterations required for convergence which is desirable in practice. On the other hand, we also investigate the performance of an AO-based scheme employing the conventional IRS model. In particular, we apply the AO-based algorithm proposed in \cite{yu2019robust} to jointly optimize the phase shifts of all IRS elements and the transmit beamformers for minimization of the BS transmit power under the same QoS requirements of the ERs and IRs as in this paper. For fair comparison, we assume that the number of scatterers in this AO-based scheme is identical to that in the proposed schemes. As can be seen from Figure \ref{poweriteraation_SWIPT}, for a large IRS with $600$ elements, the AO-based scheme employing the conventional IRS model requires more than $400$ iterations to converge which is substantially more than the proposed suboptimal scheme. This is due to the fact that by adopting an element-wise optimization framework, the search space of the AO-based scheme scales with the large number of phase shift elements. Meanwhile, we also observe that the AO-based scheme employing the conventional IRS model leads to a higher power consumption (roughly $1$ dB) compared to the proposed schemes with $T=3$ and $S=8$. The reason behind this is two-fold. First, the conventional IRS model over-optimistically assumes a unit gain for the reflected signal, which neither takes into account the limited number of scatterers nor captures the impact of the physical characteristics of the IRS channels, e.g., the incident and reflected angles and the polarization of the waves. As a result, the conventional IRS model may not be able to efficiently enhance the propagation conditions of practical low-rank channels induced by a limited number of scatterers. Second, the AO-based algorithm converges only to a stationary point \cite{bezdek2002some} while our proposed suboptimal scheme yields a close-to-optimal performance. For reference, we also show the performance of a conventional SWIPT system without IRS. In this case, the beamforming vectors and the covariance matrix of the energy signal are jointly designed for minimization of the total transmit power. As can be seen from the figure, this scheme requires a significantly higher power compared to the proposed schemes (roughly $16$ dB). This indicates that IRSs are indeed a powerful tool to enhance the performance of SWIPT systems.
\subsection{Total Transmit Power versus Minimum Required SINR}
%%%%%%%%%%%%%%%%%%%%%%%%%%%%%%%%%%%%
In Figure \ref{powerSINR_SWIPT}, we investigate the average total transmit power versus the minimum required SINR, $\Gamma_{\mathrm{req}}=\Gamma_{\mathrm{req}_k}$, $\forall k$, at the IRs for different resource allocation schemes. Since the AO-based scheme employing the conventional IRS model has a prohibitively high complexity for the considered large IRS and worse performance compared to our proposed schemes, we do not consider it in Figure \ref{powerSINR_SWIPT}. We can observe from Figure \ref{powerSINR_SWIPT} that the required total transmit powers of the proposed optimal and suboptimal schemes as well as the three baseline schemes grow with $\Gamma_{\mathrm{req}}$. This is attributed to the fact that to satisfy a more stringent minimum SINR requirement, $\Gamma_{\mathrm{req}}$, the BS has to transmit with higher power. However, the proposed optimal and suboptimal schemes achieve significant power savings compared with the three baseline schemes. This reveals the effectiveness of the proposed schemes in jointly optimizing the beamforming vectors and the transmission mode selection. Besides, we observe that for a smaller $E_{\mathrm{req}}=E_{\mathrm{req}_j}$, $\forall j$, the total transmit power for the proposed optimal and suboptimal schemes decreases. This is due to the fact that the BS has to allocate less power to the energy signal when the minimum EH requirement is less stringent. Moreover, we also show results for a scheme that is based on the overly-simplified linear EH model for the ERs. In particular, in this case, we solve a problem similar to \eqref{prob1} except that the harvested energy is assumed to be linearly proportional to the received RF power. Then, we take the obtained solution back into the actual system with non-linear EH and check if the QoS requirement of the ERs is satisfied. If the obtained solution is infeasible, we increase the transmit power until constraint C2 in \eqref{prob1} is fulfilled. As can be observed from Figure \ref{powerSINR_SWIPT}, to satisfy the QoS requirement of the ERs, the scheme based on the linear EH model consumes more power than the proposed schemes which are based on the non-linear EH model. This is due to the fact that in systems with practical non-linear EH circuits, the beamforming policy optimized for the linear EH model causes some mismatch and underutilization of resources.%However, as we further increase the size of the transmission mode set to $300$, the additional power savings of the proposed scheme are limited. In fact, the proposed scheme with $S=300$ transmission modes closely approaches the performance of the AO-based scheme employing the conventional phase shift model where all the IRS elements are jointly optimized. We note that, for an even larger IRS (e.g. more than $1000$ phase shift elements), the AO-based scheme with the conventional IRS model would be prohibitively complex, while for the proposed scheme the values of $T$ and $S$  can still be properly set for efficient online optimization. Furthermore, we also study the transmit power of the proposed scheme for different transmission mode set sizes.  In particular, as the size of the transmission mode set increases from $50$ to $100$, the BS consumes less power to meet the QoS requirements of the IRs and ERs. This is due to the fact that for a larger transmission mode set size, the IRS can perform more precise beamforming which results in power savings.
%%%%%%%%%%%%%%%%%%%%%%%%%%%%%%%%%%%%%%%%
%%%%%%%%%%%%%%%%%%%%%%%%%%%%%%%%%%%%%%%%%%%%%%%%%%%%%%%%
\begin{figure}[t]
\centering
%\vspace*{-2mm}
\hspace*{-4mm}\begin{minipage}[b]{0.47\linewidth}
    \centering
\hspace*{-2mm}\includegraphics[width=3.3in]{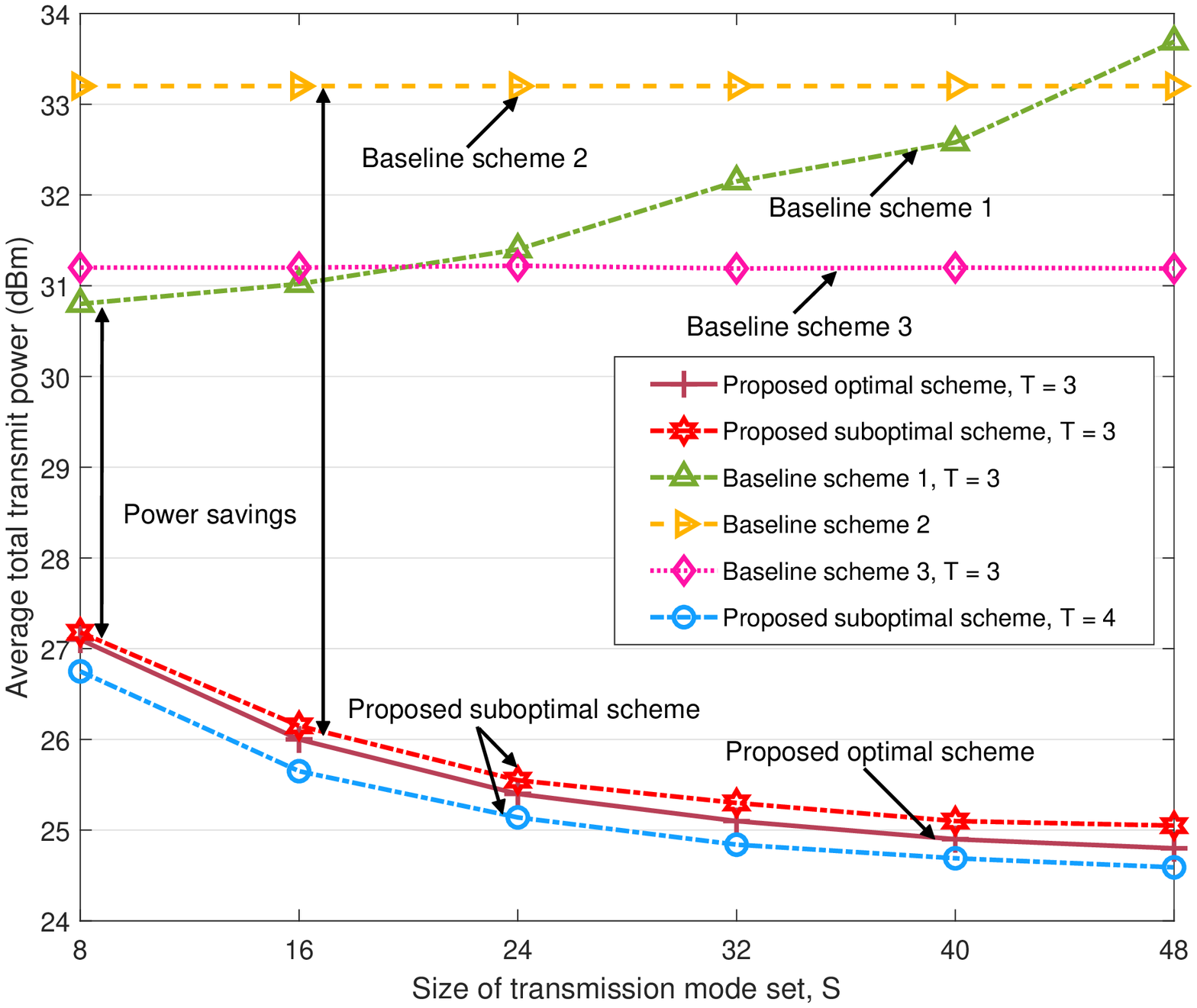}
\vspace*{-6mm}
\caption{Average total transmit power (dBm) versus the size of the refined transmission mode set for different schemes with $K=J=2$, $N_{\mathrm{T}}=8$, $\Gamma_{\mathrm{req}}=10$ dB, and $E_{\mathrm{req}}=10$ $\mu$W.}\label{powerSize_SWIPT}
\end{minipage}
\hspace*{4mm}\begin{minipage}[b]{0.47\linewidth}
    \centering
\hspace*{-2mm}\includegraphics[width=3.3in]{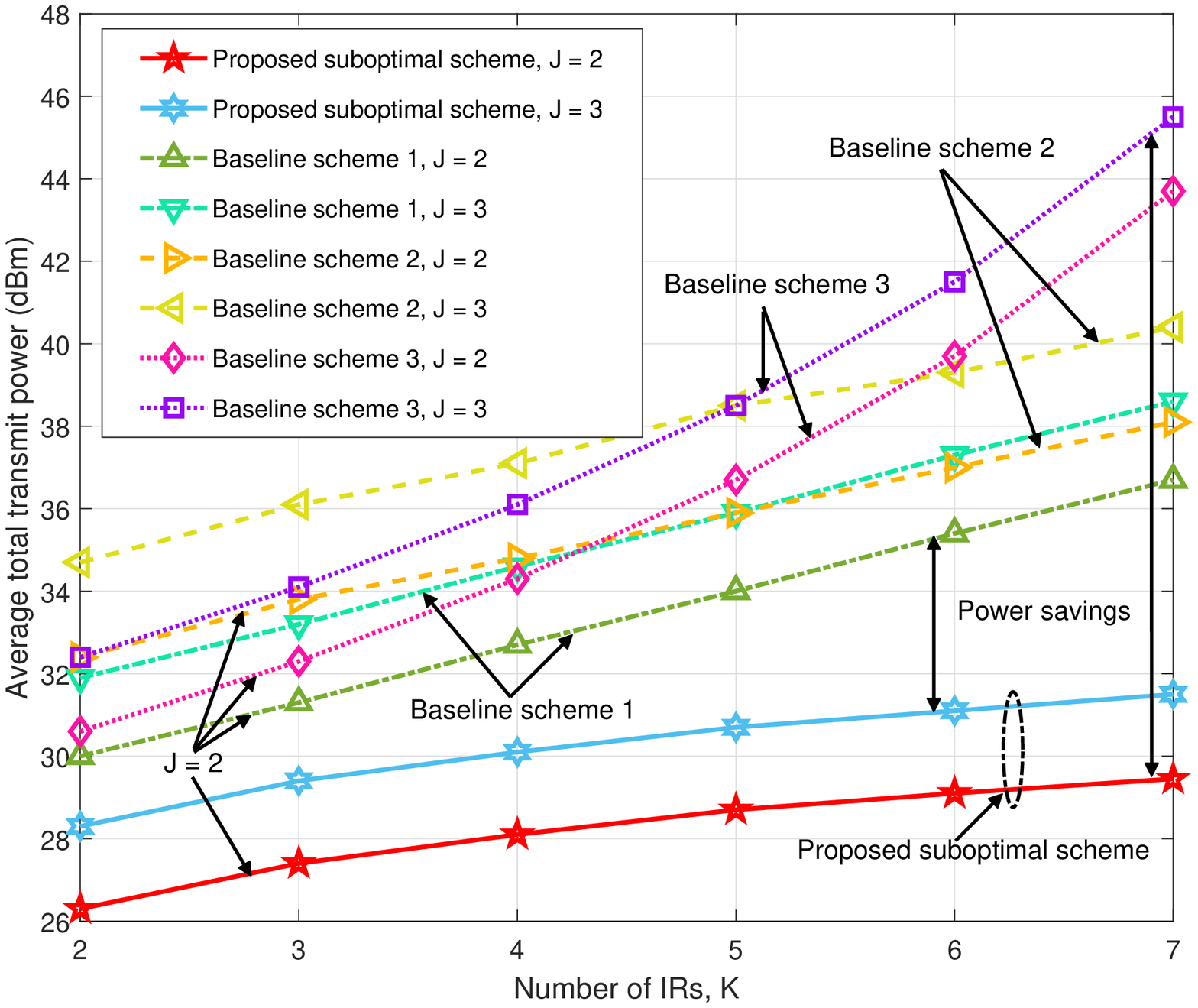}
\vspace*{-6mm}
\caption{Average total transmit power (dBm) versus the number of IRs for different schemes with $N_{\mathrm{T}}=10$, $T=3$, $S=8$, $\Gamma_{\mathrm{req}}=10$ dB, and $E_{\mathrm{req}}=10$ $\mu$W.}\label{powerReceivers_SWIPT}
\end{minipage}\vspace*{-8mm}
\end{figure}
%%%%%%%%%%%%%%%%%%%%%%%%%%%%%%%%%%%%%%%%%%%%%%%%%%
%\begin{figure}[t]\vspace*{0mm}
 %\centering
%\includegraphics[width=3.8in]{Power_Receivers_journal.eps}
%\vspace*{-4mm}
%\caption{Average total transmit power (dBm) versus the number of IRs for different resource allocation schemes with $K=2$, $J=2$, $N_{\mathrm{T}}=10$, $S=16$, $M_t=200$, $T=3$, $\Gamma_{\mathrm{req}}=10$ dB, and $E_{\mathrm{req}_j}=10$ $\mu$W.}\vspace*{-2mm}\label{powerReceivers_SWIPT}
%\end{figure}
%%%%%%%%%%%%%%%%%%%%%%%%%%%%%%%%%%%%%%%%%%%%%%%%%%%%%%%%%%%%%%%
%\begin{figure}[t]\vspace*{0mm}
 %\centering
%\includegraphics[width=3.8in]{Power_Size_journal.eps}
%\vspace*{-4mm}
%\caption{Average total transmit power (dBm) versus the size of the transmission mode set for different resource allocation schemes with $K=2$, $J=2$, $N_{\mathrm{T}}=8$, $S=8$, $M_t=200$, $T=3$, $\Gamma_{\mathrm{req}}=10$ dB, and $E_{\mathrm{req}_j}=10$ $\mu$W.}\vspace*{-2mm}\label{powerSize_SWIPT}
%\end{figure}
%%%%%%%%%%%%%%%%%%%%%%%%%%%%%%%%%%%%
\subsection{Total Transmit Power versus Size of Refined Transmission Mode Set}
Figure \ref{powerSize_SWIPT} depicts the average total transmit power versus the size of the refined transmission mode set, $S$, for different resource allocation schemes. By adjusting parameter $\delta_1$ in Criterion 1, we select more transmission modes and increase the size of the refined transmission mode set as desired.
As can be seen from the figure, the average total transmit powers of the proposed optimal and suboptimal schemes decrease with the size of the refined transmission mode set, i.e., $S$. 
This is due to the fact that as $S$ grows, additional transmission modes are included in the refined transmission mode set, which can be exploited for customizing a more favorable wireless channel and to potentially reduce the BS transmit power at the expense of a higher computational complexity. 
We note that for practical IRSs (usually comprising more than $500$ phase shift elements), algorithms developed under the conventional element-wise optimization framework, e.g., AO and IA, become prohibitively complex, while the values of $T$ and $S$ for the proposed scheme can still be properly chosen to allow for efficient online optimization. 
Moreover, unlike the proposed optimal and suboptimal schemes, the average total transmit power of baseline scheme 1 dramatically increases with $S$. 
In fact, as the refined transmission mode set becomes larger, the random transmission mode selection in baseline scheme 1 is more likely to choose a transmission mode yielding a small effective channel gain, which potentially degrades the received power of the desired signal. 
As a result, the BS has to consume more power to satisfy the QoS requirements of the receivers. In contrast, the average total transmit powers of baseline schemes 2 and 3 are almost independent of $S$. Yet, the reasons behind this are rather different. 
In particular, instead of selecting a pre-defined transmission mode, baseline scheme 2 applies an IRS with randomly generated phase shifts. Hence, the performance of baseline scheme 2 does not depend on the size of the refined transmission mode set. 
As for baseline scheme 3, the transmission mode selection strategy is based on the offline transmission mode set and identical for all tiles regardless of the size of the refined transmission mode set.
%%%%%%%%%%%%%%%%%%%%%%%%%%%%%%%%%%%
\subsection{Total Transmit Power versus Number of Receivers}
In Figure \ref{powerReceivers_SWIPT}, we study the average total transmit power versus the number of IRs, $K$, for different resource allocation schemes. 
For ease of presentation, we focus on the proposed suboptimal scheme as it closely approaches the performance of the proposed optimal scheme but entails a much lower computational complexity. 
As expected, the total transmit power increases with the number of receivers. The reason for this is two-fold. 
First, to meet the additional minimum SINR and EH requirements introduced by the additional receivers, a higher transmit power at the BS is necessary. 
Secondly, as the number of receivers increases, the BS has to dedicate more degrees of freedom (DoFs) to effectively managing the more severe multiuser interference such that the BS is less capable of reducing the total transmit power. 
Moreover, we observe that the average total transmit powers for all considered baseline schemes are substantially higher than that of the proposed suboptimal scheme. 
In particular, baseline scheme 1 yields a much higher power consumption compared to the proposed scheme due to the randomly selected transmission mode and the fixed energy signal radiation pattern. As for baseline scheme 2, due to the random phase shift pattern of the IRS, the DoFs offered by the IRS cannot be fully exploited for establishing a beneficial radio propagation environment to facilitate power-efficient resource allocation. As for baseline scheme 3, since the transmission modes for all tiles are identical and the beamforming policy is partially fixed, performance is sacrificed in exchange for a simpler implementation.
%%%%%%%%%%%%%%%%%%%%%%%%%%%%%%%%%%%%%%%%%%%%%%%%%%%%%%%%%%%%%%%
%%%%%%%%%%%%%%%%%%%%%%%%%%%%%%%%%%%%%%%%%%%%%%%%%%%%%%%%%%%%%%%
\begin{figure}[t]
\centering
%\vspace*{-2mm}
\hspace*{-4mm}\begin{minipage}[b]{0.47\linewidth}
    \centering
\hspace*{-2mm}\includegraphics[width=3.3in]{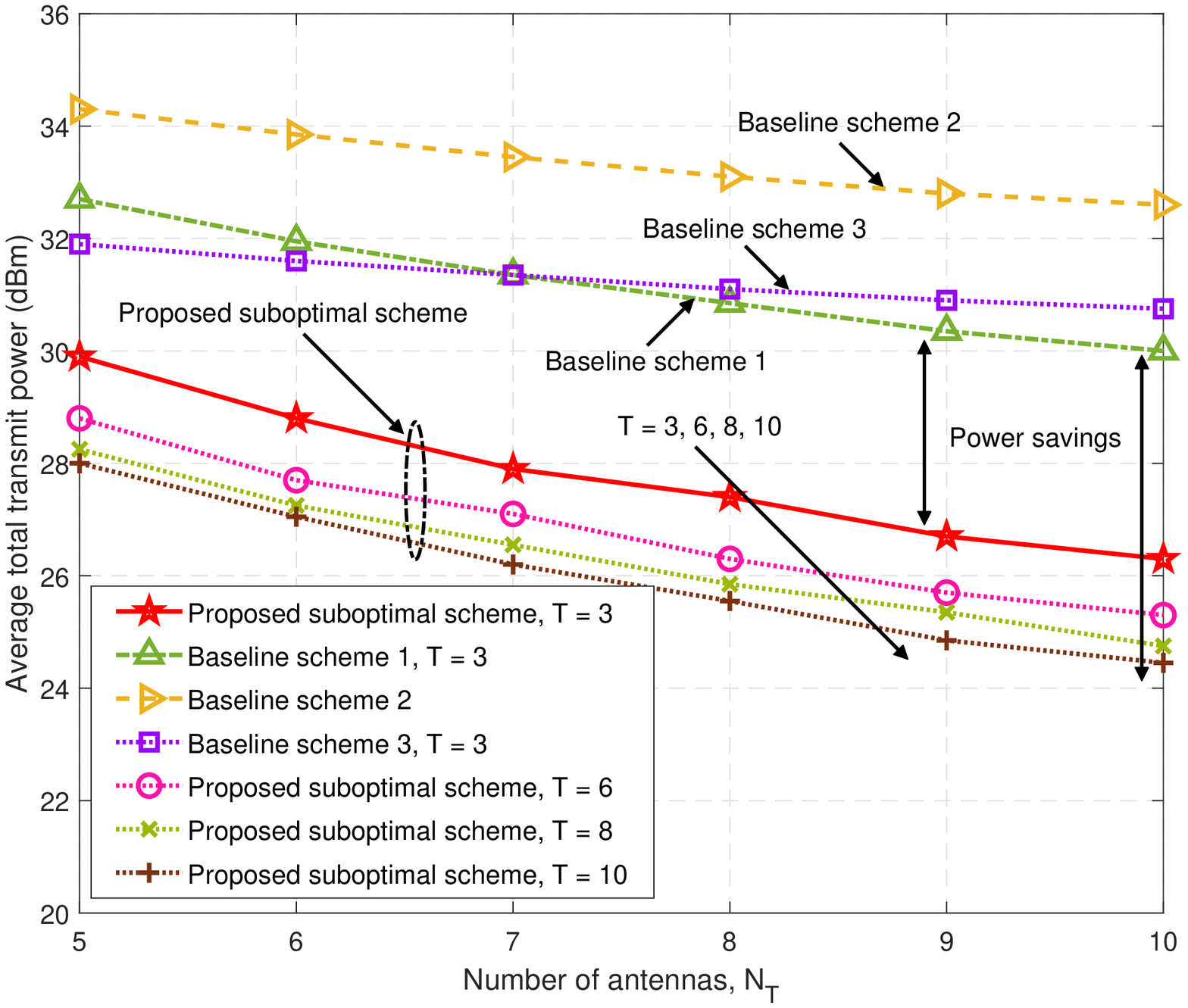}
\vspace*{-6mm}
\caption{Average total transmit power (dBm) versus the number of antennas for different schemes with $K=J=2$, $T=3$, $S=8$, $\Gamma_{\mathrm{req}_k}=10$ dB, and $E_{\mathrm{req}_j}=10$ $\mu$W.}\label{powerAntennas_SWIPT}
\end{minipage}
\hspace*{4mm}\begin{minipage}[b]{0.47\linewidth}
    \centering 
\hspace*{-2mm}\includegraphics[width=3.3in]{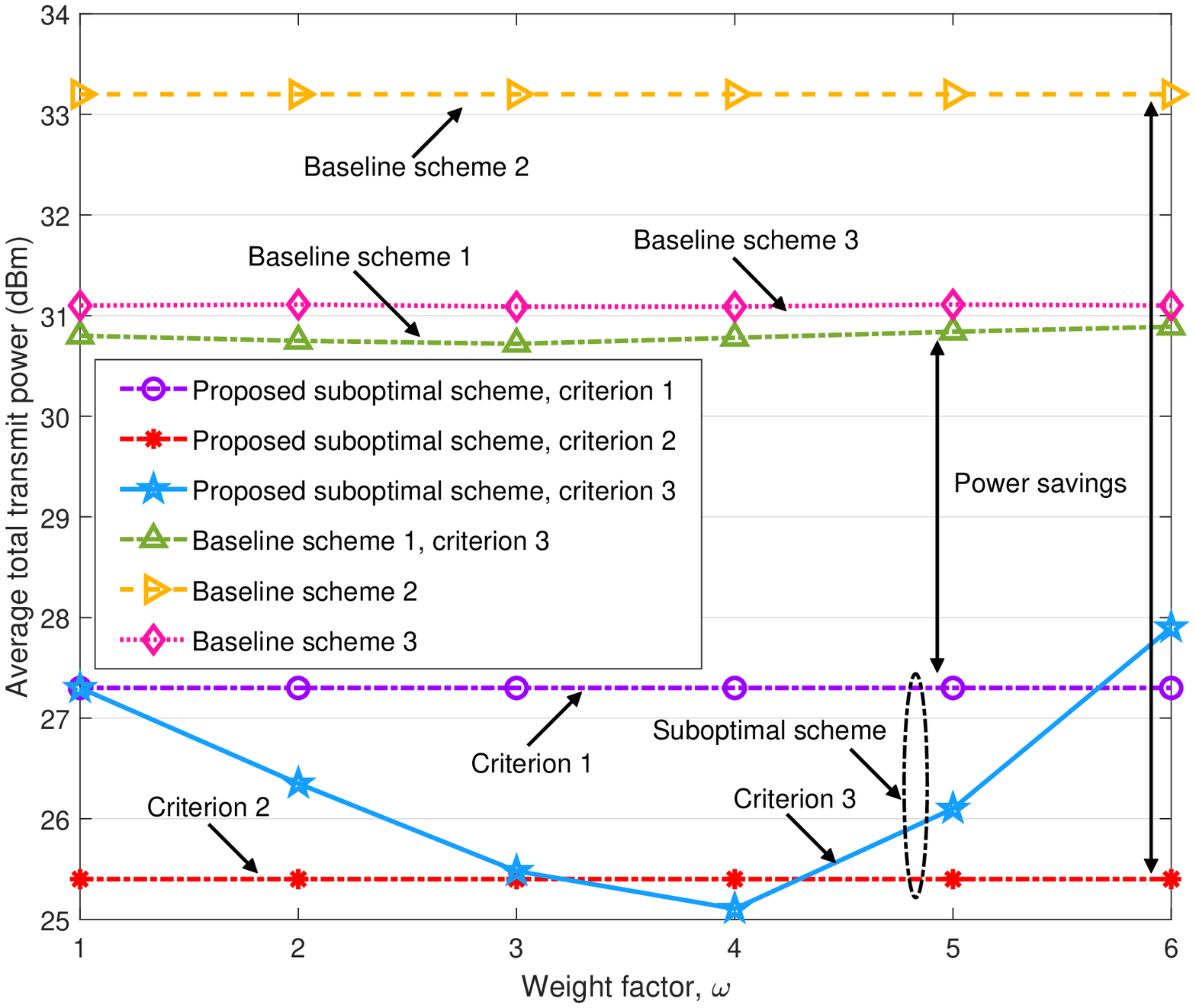}
\vspace*{-6mm}
\caption{Average total transmit power (dBm) versus the weight factor of Criterion 3 for different schemes with $K=J=2$, $N_{\mathrm{T}}=8$, $T=3$, $S=8$, $\Gamma_{\mathrm{req}}=10$ dB, and $E_{\mathrm{req}_j}=10$ $\mu$W.}\label{powerWeight_SWIPT}
\end{minipage}\vspace*{-8mm}
\end{figure}
%%%%%%%%%%%%%%%%%%%%%%%%%%%%%%%%%%%%%%%%
%%%%%%%%%%%%%%%%%%%%%%%%%%%%%%%%%%%%
\subsection{Total Transmit Power versus Number of Antennas}
Figure \ref{powerAntennas_SWIPT} illustrates the average total transmit power versus the number of antennas at the BS, $N_{\mathrm{T}}$, for different resource allocation schemes. 
It is expected that the average total transmit power decreases as the number of antennas grows since additional DoFs can be exploited for beamforming design when more antennas are available at the BS. 
Moreover, compared with the three baseline schemes, the proposed suboptimal scheme provides substantial power savings due to its ability to fully utilize the resources available in the system. On the other hand, we also study the impact of the numbers of tiles on performance. 
As can be seen from Figure \ref{powerAntennas_SWIPT}, the proposed suboptimal scheme consumes less power when the IRS is divided into more tiles. 
Yet, as we further increase $T$ from $8$ to $10$, the power reduction becomes marginal (approximately $0.2$ dB). 
In fact, a small number of tiles is sufficient to preserve most of the maximum possible performance gain enabled by the IRS. In other words, for large IRS, it is not necessary nor computationally efficient to jointly optimize all phase shift elements, as is done in the conventional element-wise optimization framework adopted in the literature. 
%Furthermore, we also investigate the performance of the schemes with different number of channel scatterers and assume $L_{\mathrm{T}}=L_{\mathrm{R}_i}=L_{\mathrm{D}_i}=L$, $\forall i\in\left\{1,\cdots,K+J\right\}$. We can observe from Figure \ref{powerAntennas_SWIPT} that the average total transmit power decreases as the number of channel scatterers in the radio propagation environment increases. This can be explained as follows: from a system perspective, the rank of the large-IRS-involved channel is usually limited by the number of channel scatterers while the additional scatterers can improve the rank conditions of the corresponding channels, yielding a richer multiplexing gain; from a design perspective, as $L$ increases, there are more AoAs and AoDs available for the IRS to exploit for selecting the desired transmission mode set for each tile. This higher flexibility can be exploited by the proposed to improve the system performance.
%%%%%%%%%%%%%%%%%%%%%%%%%%%%%%%%%%%%%%%%%%%%%%%%%%%%%%%%%%%%%%%
\subsection{Impact of Transmission Mode Pre-selection}
%%%%%%%%%%%%%%%%%%%%%%%%%%%%%%%%%%%%%%%%%%%%%%%%%%%%%%%%
%\begin{figure}[t]\vspace*{0mm}
 %\centering
%\includegraphics[width=3.3in]{Power_Weights_journal.eps}
%\vspace*{-4mm}
%\caption{Average total transmit power (dBm) versus the weight factor of criterion 3 for different resource allocation schemes with $K=J=2$, $N_{\mathrm{T}}=8$, $T=3$, $S=8$, $\Gamma_{\mathrm{req}}=10$ dB, and $E_{\mathrm{req}_j}=10$ $\mu$W.}\vspace*{-4mm}\label{powerWeight_SWIPT}
%\end{figure}
Figure \ref{powerWeight_SWIPT} illustrates the average total transmit power versus the weight factor of Criterion 3 for transmission mode pre-selection. We observe that the average total transmit power for the proposed suboptimal scheme employing Criterion 3 first decreases with weight factor $\omega$ ($1 \leq \omega \leq 4$). Specifically, as the weight factor increases, more transmission modes that favor the effective channel vectors of the ERs, i.e., $\left \| \mathbf{h}_{\mathrm{E},j,s,t} \right \|_2$, are included in the refined transmission mode set. Also, in typical SWIPT systems, the ERs usually require much higher received powers compared to the IRs. As a result, refined transmission mode sets constructed with larger $\omega$ enable the configuration of wireless propagation environments that are more favorable for the ERs, which potentially results in power savings. Yet, as we further increase $\omega$, the power consumption of the BS starts to increase. This is due to the fact that very large weight factors ($\omega > 4$) lead to the construction of severely biased transmission mode sets that are favorable for the ERs but lead to poor channel conditions for the IRs. Hence, the BS is forced to consume more power to compensate for the severe signal attenuation, which outweighs the power gain resulting from customizing favorable channels for the ERs. In contrast, though also employing Criterion 3 to refine the transmission mode set, baseline scheme 1 is not sensitive to $\omega$. This is due to the fact that its transmission mode selection policy is fixed instead of being optimized over the refined transmission mode set. On the other hand, we also show the performance of the proposed suboptimal scheme employing transmission mode pre-selection Criteria 2 and 3, which do not depend on $\omega$. In particular, Criterion 2 leads to a lower transmit power compared to Criterion 1. In fact, since Criterion 1 only focuses on the magnitude of the channel, it may construct a biased transmission mode set when one receiver enjoys much better channel conditions than the other receivers. In this case, the BS is forced to increase the transmit power to satisfy the QoS requirements of the receivers with poor channel conditions. On the contrary, Criterion 2 generates individual transmission mode sets for all receivers and promotes a wireless propagation environment that is favorable for all receivers, which potentially leads to less transmit power consumption compared with Criterion 1.
%Moreover, we note that by exploiting the characteristics of the considered system and carefully choosing the weight factor, the desired system performance may further be reduced.
%%%%%%%%%%%%%%%%%%%%%%%%%%%%%%%%%%%%%%%%%%%%%%%%%%%%%%%%%%%%%%%
\section{Conclusion}
\label{Large_IRS_conclusion}
In this paper, we studied the resource allocation algorithm design for large IRS-assisted SWIPT systems. Compared with existing works assuming an overly simplified system model, we adopted a physics-based IRS model and a non-linear EH model which can better capture the properties of practical IRS-assisted SWIPT systems. To facilitate the efficient system design for large IRSs, we partition the IRS into several tiles and adopt a TT-based optimization framework which comprises an offline transmission mode set design stage and an online optimization stage. To further reduce the computational complexity of IRS online design, we proposed two new transmission mode pre-selection criteria. Given the refined transmission mode set, we focused on the joint online optimization of the transmit beamforming vectors, the covariance matrix of the energy signal, and the transmission mode selection for minimization of the BS transmit power while satisfying the QoS requirements of the IRs and the ERs. To tackle the formulated combinatorial optimization problem, we first proposed a BnB-based optimization algorithm which yields the globally optimal solution of the considered optimization problem. Since the optimal scheme entails a high computational complexity, we also developed a computationally efficient SCA-based algorithm which asymptotically converges to a locally optimal solution. Simulation results showed that the proposed schemes do not only yield considerable power savings compared with three baseline schemes but also allow us to flexibly strike a balance between system performance and computational complexity by adjusting the number of tiles and transmission modes. Moreover, the adopted physics-based IRS model can effectively leverage the wave AoA and AoD and polarization that are not explicitly modeled by the conventional IRS model. Furthermore, in combination with the physics-based IRS model, the proposed TT-based framework was shown to be crucial for realizing real-time online design of wireless systems assisted by large IRS.
%\vspace*{-2mm}
%%%%%%%%%%%%%%%%%%%%%%%%%%%%%%%%%%%%%%%%%%%%%%%%%%%%%%%%%%%%%%
\section*{Appendix}
%\subsection{Proof of Theorem 1}
To start with, we rewrite the relaxed version of problem \eqref{prob3} equivalently as follows:
\begin{eqnarray}
\label{prob7}
&&\hspace*{-4mm}\underset{\substack{\mathbf{W}_k,\widetilde{\mathbf{W}}_{k,s,t,p,q}\in\mathbb{H}^{N_{\mathrm{T}}},\\\mathbf{V},\widetilde{\mathbf{V}}_{s,t,p,q}\in\mathbb{H}^{N_{\mathrm{T}}},\\\widetilde{b}_{s,t},\widetilde{\beta}_{ s,t,p,q},\theta_j}}{\mino} \,\, \,\, \underset{\substack{s\in\mathcal{S},t\in\widehat{\mathcal{T}}}}{\sum }\underset{\substack{p\in\mathcal{S},q\in\widehat{\mathcal{T}}}}{\sum }\mathrm{Tr}\left (\underset{k\in\mathcal{K}}{\sum }\widetilde{\mathbf{W}}_{k,s,t,p,q}+\widetilde{\mathbf{V}}_{s,t,p,q}\right )\notag\\
&&\widetilde{\mbox{C1}},\mbox{C3},\widetilde{\mbox{C4}},\widetilde{\mbox{C5}},\widetilde{\mbox{C6}}\mbox{a}\mbox{-}\widetilde{\mbox{C6}}\mbox{d},\widetilde{\mbox{C7}}\mbox{a}\mbox{-}\widetilde{\mbox{C7}}\mbox{d},\widetilde{\mbox{C8}}\mbox{a}\mbox{-}\widetilde{\mbox{C8}}\mbox{d},~\widetilde{\mbox{C2}}\mbox{a}\mbox{:~}C_{\mathrm{req}_j}\geq \mathrm{exp}\left (-\varrho _j\theta_j \right ),~\forall j,\notag\\
&&\widetilde{\mbox{C2}}\mbox{b}\mbox{:~}\theta_j\geq\underset{\substack{s\in\mathcal{S},t\in\widehat{\mathcal{T}}}}{\sum }\underset{\substack{p\in\mathcal{S},q\in\widehat{\mathcal{T}}}}{\sum }\mathrm{Tr}\Big( \mathbf{g}_{p,q,j} \mathbf{g}^H_{s,t,j}\big( \underset{k\in\mathcal{K}}{\sum }\widetilde{\mathbf{W}}_{k,s,t,p,q}+\widetilde{\mathbf{V}}_{s,t,p,q}\big)\Big),~\forall j.
\end{eqnarray}
Note that the optimization problem in \eqref{prob7} is jointly convex with respect to the optimization variables and satisfies Slater’s constraint qualification. Thus, strong duality holds for \eqref{prob7}. Moreover, since $\widetilde{\mathbf{W}}_{k,s,t,p,q}=\widetilde{\beta}_{ s,t,p,q}\mathbf{W}_k$ always holds, we express the Lagrangian function of \eqref{prob7} in terms of $\mathbf{W}_k$ as follows
\begin{eqnarray}
\label{Lagrangian}
\mathcal{L}&&\hspace*{-6mm}=\hspace*{-4mm}\hspace*{2mm}\underset{k\in\mathcal{K}}{\sum } \Big(\mathrm{Tr}(\underset{\substack{s,p\in\mathcal{S},\\q,t\in\widehat{\mathcal{T}}}}{\sum }\widetilde{\beta}_{ s,t,p,q}\mathbf{W}_k)+\mathrm{Tr}(\underset{\substack{s,p\in\mathcal{S},\\q,t\in\widehat{\mathcal{T}}}}{\sum }\widetilde{\beta}_{ s,t,p,q}\mathbf{U}_{k,s,t,p,q}\mathbf{W}_k)-\eta_k\mathrm{Tr}(\underset{\substack{s,p\in\mathcal{S},\\q,t\in\widehat{\mathcal{T}}}}{\sum }\widetilde{\beta}_{ s,t,p,q}\mathbf{h}_{\mathrm{I},k,p,q}\mathbf{h}_{\mathrm{I},k,s,t}^H\mathbf{W}_k ) \notag\\
&&\hspace*{-6mm}-\mathrm{Tr}(\underset{\substack{s,p\in\mathcal{S},\\q,t\in\widehat{\mathcal{T}}}}{\sum }\widetilde{\beta}_{ s,t,p,q}\mathbf{Z}_{k,s,t,p,q}\mathbf{W}_k)+\underset{\substack{s,p\in\mathcal{S},\\q,t\in\widehat{\mathcal{T}}}}{\sum }(1-\widetilde{\beta}_{ s,t,p,q})\big(\mathrm{Tr}(\mathbf{X}_{k,s,t,p,q}\mathbf{W}_k)-\mathrm{Tr}(\mathbf{Y}_{k,s,t,p,q}\mathbf{W}_k)\big)
\notag\\
&&\hspace*{-6mm}+\eta_k\Gamma_k^{\mathrm{req}}\underset{r\in\mathcal{K}\setminus \left \{ k \right \} }{\sum }\mathrm{Tr} (\underset{\substack{s,p\in\mathcal{S},\\q,t\in\widehat{\mathcal{T}}}}{\sum }\widetilde{\beta}_{ s,t,p,q}\mathbf{h}_{\mathrm{I},k,p,q}\mathbf{h}_{\mathrm{I},k,s,t}^H \mathbf{W}_k)+\underset{j\in\mathcal{J}}{\sum }\zeta _j\mathrm{Tr}(\underset{\substack{s,p\in\mathcal{S},\\q,t\in\widehat{\mathcal{T}}}}{\sum }\widetilde{\beta}_{ s,t,p,q}\mathbf{h}_{\mathrm{E},j,p,q}\mathbf{h}_{\mathrm{E},j,s,t}^H\mathbf{W}_k)\Big)\hspace*{-1mm}+\hspace*{-1mm}\Phi.
\end{eqnarray}
Here, $\Phi$ comprises the terms that do not involve $\mathbf{W}_k$. The scalar Lagrange multipliers $\eta_k$ and $\zeta_j\geq0$ are associated with constraint $\widetilde{\mbox{C1}}$ and $\widetilde{\mbox{C2}}\mbox{b}$, respectively. The positive semidefinite Lagrange multiplier matrices $\mathbf{U}_{k,s,t,p,q}$, $\mathbf{X}_{k,s,t,p,q}$, $\mathbf{Y}_{k,s,t,p,q}$, and $\mathbf{Z}_{k,s,t,p,q}\in \mathbb{C}^{N_\mathrm{T}\times N_\mathrm{T}}$ are associated with constraints $\widetilde{\mbox{C7}}\mbox{a}$, $\widetilde{\mbox{C7}}\mbox{b}$, $\widetilde{\mbox{C7}}\mbox{c}$, and $\widetilde{\mbox{C7}}\mbox{d}$. Note that there always exist $\widetilde{\beta}_{s,t,p,q}^*>0$. Next, by examining the KKT conditions with respect to $\mathbf{W}_k$, we investigate the structure of the optimal beamforming matrix. In particular, we have
\begin{eqnarray}
\label{KKTCond1}
&&\hspace*{-10mm}\mbox{K1}:~\eta_k^*,\zeta_j^*\geq 0,\hspace*{2mm}\mathbf{U}_{k,s,t,p,q}^*,\mathbf{X}_{k,s,t,p,q}^*,\mathbf{Y}_{k,s,t,p,q}^*,\mathbf{Z}_{k,s,t,p,q}^*\succeq \mathbf{0},\\
&&\hspace*{-10mm}\mbox{K2}:~\underset{\substack{s\in\mathcal{S},t\in\widehat{\mathcal{T}}}}{\sum }\underset{\substack{p\in\mathcal{S},q\in\widehat{\mathcal{T}}}}{\sum }\widetilde{\beta}_{s,t,p,q}^*\mathbf{Z}_{k,s,t,p,q}^*\mathbf{W}_k^*=\mathbf{0},\hspace*{20mm}\mbox{K3}:~\nabla_{\mathbf{W}_k}\mathcal{L}(\mathbf{W}_k^*)=\mathbf{0},\label{KKTCond2}
\end{eqnarray}
where $\eta_k^*$, $\zeta_j^*$, $\mathbf{U}_{k,s,t,p,q}^*$, $\mathbf{X}_{k,s,t,p,q}^*$, $\mathbf{Y}_{k,s,t,p,q}^*$, and $\mathbf{Z}_{k,s,t,p,q}^*$ are the optimal values corresponding to $\mathbf{W}_k^*$ and $\widetilde{\beta}_{s,t,p,q}^*$. Note that $\widetilde{\beta}_{s,t,p,q}^*\geq0$, $\mathbf{W}_k^*\succeq \mathbf{0}$, and $\mathbf{Z}_{k,s,t,p,q}^*\succeq \mathbf{0}$, and we have $\widetilde{\beta}_{s,t,p,q}^*\mathbf{Z}_{k,s,t,p,q}^*\mathbf{W}_k^*=\mathbf{0}$. Then, assuming $\widetilde{\beta}_{s^{\circ },t^{\circ },p^{\circ },q^{\circ }}^*=1$, we explicitly write $\nabla_{\mathbf{W}_k}\mathcal{L}(\mathbf{W}_k^*)$ in $\mbox{K3}$ as follows
\begin{equation}
\label{Large_IRS_largrargian_multiplier}
    \mathbf{Z}_{k,s^{\circ },t^{\circ },p^{\circ },q^{\circ }}^*=\underset{\substack{s\in\mathcal{S},t\in\widehat{\mathcal{T}}}}{\sum }\underset{\substack{p\in\mathcal{S},q\in\widehat{\mathcal{T}}}}{\sum }\widetilde{\beta}_{s,t,p,q}^*\mathbf{I}_{N_{\mathrm{T}}}-\mathbf{\Delta}_{k}^*,
\end{equation}
where $\mathbf{\Delta}_{k}^*$ is given by
\begin{eqnarray}
\mathbf{\Delta}_{k}^*&&\hspace*{-6mm}=\underset{\substack{s,p\in\mathcal{S},\\q,t\in\widehat{\mathcal{T}}}}{\sum }\Big(\widetilde{\beta}_{s,t,p,q}^*\eta_k^*\mathbf{h}_{\mathrm{I},k,p,q}\mathbf{h}_{\mathrm{I},k,s,t}^H-\widetilde{\beta}_{s,t,p,q}^*\hspace*{-1mm}\underset{r\in\mathcal{K}\setminus \left \{ k \right \} }{\sum }\eta_r^*\Gamma_r^{\mathrm{req}}\mathbf{h}_{\mathrm{I},r,p,q}\mathbf{h}_{\mathrm{I},r,s,t}^H-\widetilde{\beta}_{s,t,p,q}^*\underset{j\in\mathcal{J}}{\sum }\zeta _j^* \mathbf{h}_{\mathrm{E},j,p,q}\mathbf{h}_{\mathrm{E},j,s,t}^H\notag\\
&&\hspace*{-6mm}-\widetilde{\beta}_{s,t,p,q}^*\mathbf{U}_{k,s,t,p,q}^*+(1-\widetilde{\beta}_{s,t,p,q}^*)\big(\mathbf{Y}_{k,s,t,p,q}^*-\mathbf{X}_{k,s,t,p,q}^*\big)\Big)+\underset{\substack{s\in\mathcal{S}\setminus \left \{ s^{\circ } \right \},\\t\in\widehat{\mathcal{T}}\setminus \left \{ t^{\circ } \right \}}}{\sum }\underset{\substack{p\in\mathcal{S}\setminus \left \{ p^{\circ } \right \},\\q\in\widehat{\mathcal{T}}\setminus \left \{ q^{\circ } \right \}}}{\sum }\widetilde{\beta}_{s,t,p,q}^*\mathbf{Z}_{k,s,t,p,q}^*.
\end{eqnarray}
Then, using similar arguments as in \cite[Appendix A]{xu2020resource}, it can be shown that $\mathbf{Z}_{k,s^{\circ },t^{\circ },p^{\circ },q^{\circ }}^*$ in \eqref{Large_IRS_largrargian_multiplier} satisfies $\mathrm{Rank}(\mathbf{Z}_{k,s^{\circ },t^{\circ },p^{\circ },q^{\circ }}^*)\geq N_{\mathrm{T}}-1$. Recalling $\mathbf{Z}_{k,s^{\circ },t^{\circ },p^{\circ },q^{\circ }}^*\mathbf{W}_k^*=\mathbf{0}$, for each IR $k$, we can always obtain an optimal $\mathbf{W}_k^*$ with a unit rank. This completes the proof. 
\vspace*{-1mm}
\bibliographystyle{IEEEtran}
\bibliography{Reference_List}
\end{document}